\documentclass[aps,prd,groupedaddress,nofootinbib]{revtex4}
\usepackage{graphicx,color,amsmath,mathtools,amssymb}
\usepackage{tcolorbox}
\usepackage{amsthm}
\usepackage{graphicx}
\usepackage{color,soul}
\usepackage{soul}
\usepackage{bm}
\usepackage{subfigure}
\usepackage{multirow}
\usepackage{hyperref} % For hyperlinks in the PDF
\hypersetup{colorlinks=true,citecolor=blue,linkcolor=magenta,filecolor=magenta,urlcolor=blue,}

\begin{document}
    \title{Double-charm and hidden-charm hexaquark states under the complex scaling method}

    \author{Jian-Bo Cheng$^{1}$}%
    \email{jbcheng@pku.edu.cn}
    \author{Du-xin Zheng$^{2}$}%
    \email{duxin.zheng@iat.cn}
    \author{Zi-Yang Lin$^{1}$}
    \email{lzy\_15@pku.edu.cn}
    \author{Shi-Lin Zhu$^{1}$}
    \email{zhusl@pku.edu.cn}
    \affiliation{
        $^1$School of Physics and Center of High Energy Physics, Peking University 100871, China\\
        $^{2}$Shandong Institute of Advanced Technology, Jinan 250100, China
    }%

    \date{\today}% It is always \today, today,
    % but any date may be explicitly specified

\begin{abstract}

We investigate the double-charm and hidden-charm hexaquarks as
molecules in the framework of the one-boson-exchange potential
model. The multichannel coupling and $S-D$ wave mixing are taken
into account carefully. We adopt the complex scaling method to
investigate the possible quasibound states, whose widths are from
the three-body decay channel $\Lambda_c\Lambda_c\pi$ or
$\Lambda_c\bar{\Lambda}_c\pi$. For the double-charm system of $I(J^P)=1(1^+)$,
we obtain a quasibound state, whose width is 0.50 MeV if the binding
energy is -14.27 MeV. And the $S$-wave $\Lambda_c\Sigma_c$ and
$\Lambda_c\Sigma_c^*$ components give the dominant contributions.
For the $1(0^+)$ double-charm hexaquark system, we do not find any
pole. We find more poles in the hidden-charm hexaquark system. We
obtain one pole as a quasibound state in the
$I^G(J^{PC})=1^+(0^{--})$ system, which only has one channel
$(\Lambda_c\bar{\Sigma}_c+\Sigma_c\bar{\Lambda}_c)/\sqrt{2}$. Its
width is 1.72 MeV with a binding energy of -5.37 MeV. But, we do not
find any pole for the scalar $1^-(0^{-+})$ system. For the vector
$1^-(1^{-+})$ system, we find a quasibound state. Its energies,
widths and constituents are very similar to those of the $1(1^+)$
double-charm case. In the vector $1^+(1^{--})$ system, we get two
poles---a quasibound state and a resonance. The quasibound state has
a width of 0.6 MeV with a binding energy of -15.37 MeV. For the
resonance, its width is 2.72 MeV with an energy of 63.55 MeV
relative to the $\Lambda_c\bar{\Sigma}_c$ threshold. And its partial
width from the two-body decay channel
$(\Lambda_c\bar{\Sigma}_c-\Sigma_c\bar{\Lambda}_c)/\sqrt{2}$ is
apparently larger than the partial width from the three-body decay
channel $\Lambda_c\bar{\Lambda}_c\pi$. Especially, the $1^+(0^{--})$
and $1^-(1^{-+})$ hidden-charm hexaquark molecular states are very
interesting. These isovector mesons have exotic $J^{PC}$ quantum
numbers which are not accessible to the conventional $q\bar q$
mesons.

\end{abstract}
    %\pacs{14.40.Rt, 12.39.Pn}
    \maketitle
    %%%%%%%%%%%%%%%%%%%%%%%%%%%%%%%%%%%%%%%%%%%%%%%%%%%%%%
    \section{Introduction}\label{sec:Introduction}
    %%%%%%%%%%%%%%%%%%%%%%\href{}{}%%%%%%%%%%%%%%%%%%%%%%%%%%%%%%%%

In the study of the hadronic molecular states, the dibaryon always
plays an important role. The well-known deuteron is the only
experimentally confirmed baryon-baryon bound state without charm
quarks. Moreover, the WASA-at-COSY Collaboration Collaboration
repeatedly observed the dibaryon resonance $d^*(2380)$
\cite{Adlarson2011,Adlarson2013,Adlarson2013a,Adlarson2014,Bashkanov2009}.
It is natural to extend the investigation from the deuteron to the
strange dibaryon. Jaffe suggested the famous H-dibaryon composed of
the $\Lambda\Lambda$ pair \cite{Jaffe1977}, which was also studied
in a series of works
\cite{Jaffe1977,Li2018,Morita2015,Inoue2011,Beane2011,Yoon2007,Polinder2007,Takahashi2001,Paganis2000,Karl1987,Rosner1986,Yost1985,Mackenzie1985,Balachandran1984,Balachandran1984a}.
Besides, the dibaryon with one heavy quark ($qqqqqQ$) was also
investigated in Refs. \cite{Fromel2005,Liu2012,Huang2013}  .

In 2017, the LHCb Collaboration discovered a double-charm baryon in
the $\Lambda_{c}^{+}K^-\pi^+\pi^-$ mass spectrum---$\Xi_{cc}^{++}$
\cite{Aaij2017}, which is also the first observed double-heavy
hadron. This discovery encouraged the research on the double-heavy
hadrons, especially the double-charm tetraquarks
\cite{luoExoticTetraquarkStates2017,mehenImplicationsHeavyQuarkdiquark2017a,fontouraProductionExoticTetraquarks2019a,xuPotentialsChiralEffective2019,francisEvidenceCharmbottomTetraquarks2019a,agaevStrongDecaysDoublecharmed2019a,tanSystematicsQQBarq2020,yangDoubleheavyTetraquarks2020,chengDoubleheavyTetraquarkStates2021}.
After four years, the LHCb Collaboration reported the observation of
the first double-heavy exotic hadron---$T_{cc}^{+}$
\cite{aaijStudyLineshape38722020,Aaij2022}. After the rapid
succession of the double-charm hadron discovery, it is necessary to
implement a further investigation. We will focus on the double-charm
deuteron-like hexaquarks in this work.

In the molecule picture, the double-charm dibaryon could be easier
to form a bound state due to the larger reduced mass. And in fact,
the  double-heavy hexaquark ($qqqqQQ$) systems have been discussed
in various approaches \cite{Carames2015,Garcilazo2020,Huang2014,Xia2022,Liu2022b,Wang2021a,Chen2022,Lu2019,Chen2017,Lee2011,Li2012,Ling2021,Meguro2011,Meng2018,Yu2021,Vijande2016,Liu2022a,Dong2021e,Gerasyuta2012},
including the chiral constituent quark model
\cite{Carames2015,Garcilazo2020}, the quark delocalization color
screening model \cite{Huang2014,Xia2022}, the chromomagnetic model
\cite{Liu2022b}, the QCD sum rules \cite{Wang2021a}, the chiral
effective field theory (EFT) \cite{Chen2022,Lu2019}, and the
one-boson-exchange (OBE) model
\cite{Chen2017,Lee2011,Li2012,Ling2021,Meguro2011,Meng2018,Yu2021}.
For the $\Lambda_c\Sigma_c$ molecule system, the authors of Ref.
\cite{Li2012} found a bound state with the number $I(J^P)=1(1^+)$.
However, it was not confirmed in Ref. \cite{Xia2022}. The
hidden-heavy hexaquark $(qqQ\bar{q}\bar{q}\bar{Q})$ may have similar
behaviors, and the relevant investigations can be found in Refs.
\cite{Chen2012,Chen2013,Chen2016,Liu2022a,Qiao2006,Qiao2008,Wan2020}.

In this work, we investigate the double-charm dibaryon and
hidden-charm baryonium systems containing the
$\Lambda_c\Sigma_c^{(*)}$ and $\Lambda_c\bar{\Sigma}_c^{(*)}$
channels in the molecule picture. As pointed out in our previous
work \cite{Cheng2022b}, the cross diagram $DD^{*}\leftrightarrow
D^{*}D$ of the one-pion-exchange will provide a complex potential,
which is from the three-body decay effect. This behavior could also
occur in the process $\Lambda_c\Sigma_c^{(*)}\leftrightarrow
\Sigma_c^{(*)}\Lambda_c$. To study the possible three-body effect,
we will retain the imaginary contribution from the OPE potential.

We use the OBE model to deal with the molecule state system. In
order to explore the existence of the resonance, we will adopt the
complex scaling method (CSM)
\cite{aguilarClassAnalyticPerturbations1971,balslevSpectralPropertiesManybody1971b},
which is a powerful method that can handle the bound state and
resonance in a consistent way. Besides, the $S-D$ wave mixing and
coupled-channel effect will be considered in this work. For both the
double-charm and hidden-charm hexaquarks, the dominant contributions
of the widths arise from the open-charm decay processes. The
possible hidden-charm decay contributions for the hidden-charm
hexaquark systems are negligible.

This paper is organized as follows. In Sec. \ref{sec:framework}, we
will introduce our framework explicitly. In Sec. \ref{sec:
potentials}, we present the effective Lagrangians and potentials. In
Sec. \ref{sec: results}, we solve the complex scaled Schr\"odinger
equation and give the results of the double-charm dibaryon and
hidden-charm baryonium. The last section \ref{sec:summary} is a
brief summary.

%%%%%%%%%%%%%%%%%%%%%%%%%%%%%%%%%%%%%%%%%%%%%%%%%%%%%%
\section{Framework}\label{sec:framework}

In the previous work \cite{Cheng2022b}, we studied the double-charm
tetraquark with the CSM method. The $DD^*$ system is found to be
special since the 0th component of the transferred momentum of the
exchanged pion is larger than the pion mass. This feature will
provide the OPE potential with an imaginary part. If one could get a
pole in this system, one may get an energy with an imaginary part
which is explained as its half width. Therefore, we will pay more
attention to the systems containing this type of interaction. One
could see similar interactions in several systems, such as
$\Lambda_c\Sigma_c\leftrightarrow\Sigma_c\Lambda_c$,
$\Lambda_c\bar{\Sigma}_c\leftrightarrow\Sigma_c\bar{\Lambda}_c$,
$\Lambda_cD^*\leftrightarrow\Sigma_cD$ and
$\Lambda_c\bar{D}^*\leftrightarrow\Sigma_c\bar{D}$. In this work, we
consider the former two cases: the double-charm and hidden-charm
hexaquark molecule system.

To find the possible bound and resonant states, we take into account
coupled-channel effect. The explicit systems and channels can be seen
in Table \ref{tab: channel}. However, we do not consider the
isoscalar system with channels $\Lambda_c\Lambda_c$,
$\Sigma_c\Sigma_c$ and $\Sigma_c^*\Sigma_c^*$ (or
$\Lambda_c\bar{\Lambda}_c$, $\Sigma_c\bar{\Sigma}_c$ and
$\Sigma_c^*\bar{\Sigma}_c^*$) in this work. Unlike the other two
isovector systems, this system does not have the special cross
diagram and could not contribute an imaginary part to the OPE
potential. We will study these systems in the subsequent work. For
the isovector cases, we will not take into account channels
$\Sigma_c\Sigma_c^*$ and $\Sigma_c^*\Sigma_c^*$ (or
$\Sigma_c\bar{\Sigma}_c^*$ and $\Sigma_c^*\bar{\Sigma}_c^*$) due to
the same reason. In this work, we only consider the channels with
$^1S_0$ ($J$=0), $^3S_1$ ($J$=1) and $^3D_1$ ($J$=1).

The masses of the charmed baryon and exchanged light mesons are
shown in Table \ref{tab: mass meson}. We take the isospin mean
masses to deal with the isospin conservation process.

\begin{table}[htbp]
    \renewcommand{\arraystretch}{1.8}{
        \setlength\tabcolsep{8pt}{
            \begin{tabular}{c|ccccccc}\hline\hline
                \centering
                &   $I^G(J^{PC})$&$1$&$2$&$3$&$4$&$5$&$6$\\\hline
                \multirow{3}{*}{$cc$}&  $0(0^+)$&$\Lambda_c \Lambda_c(^1S_0)$&$\Sigma_c\Sigma_c(^1S_0)$&$\Sigma_c^*\Sigma_c^*(^1S_0)$&&&\\
                &    $1(0^{+})$&$\Lambda_c\Sigma_c(^1S_0)$&&&&&\\
                &    $1(1^{+})$&$\Lambda_c\Sigma_c(^3S_1)$&$\Lambda_c\Sigma_c(^3D_1)$&$\Lambda_c\Sigma_c^*(^3S_1)$&$\Lambda_c\Sigma_c^*(^3D_1)$&$\Sigma_c\Sigma_c (^3S_1)$&$\Sigma_c\Sigma_c (^3D_1)$\\
                \hline
                \multirow{5}{*}{$c\bar{c}$}&    $0^+(0^{-+})$&$\Lambda_c\bar{\Lambda}_c(^1S_0)$&$\Sigma_c\bar{\Sigma}_c(^1S_0)$&$\Sigma_c^*\bar{\Sigma}_c^*(^1S_0)$&&&\\
                &    $1^+(0^{--})$&$\{\Lambda_c\bar{\Sigma}_c\}(^1S_0)$&&&&&\\
                &    $1^-(0^{-+})$&$[\Lambda_c\bar{\Sigma}_c](^1S_0)$&$\Sigma_c\bar{\Sigma}_c(^1S_0)$&&&&\\
                & $1^+(1^{--})$&$[\Lambda_c\bar{\Sigma}_c](^3S_1)$&$[\Lambda_c\bar{\Sigma}_c](^3D_1)$&$\{\Lambda_c\bar{\Sigma}_c^*\}(^3S_1)$&$\{\Lambda_c\bar{\Sigma}_c^*\}(^3D_1)$&$\Sigma_c\bar{\Sigma}_c (^3S_1)$&$\Sigma_c\bar{\Sigma}_c (^3D_1)$\\
                & $1^-(1^{-+})$&$\{\Lambda_c\bar{\Sigma}_c\}(^3S_1)$&$\{\Lambda_c\bar{\Sigma}_c\}(^3D_1)$&$[\Lambda_c\bar{\Sigma}_c^*](^3S_1)$&$[\Lambda_c\bar{\Sigma}_c^*](^3D_1)$&&\\
                \hline\hline
    \end{tabular}}}
\caption{The channels of the double- and hidden-charm hexaquark
systems. We adopt the following shorthand notations for simplicity,
$[\Lambda_c\bar{\Sigma}_c^{(*)}]=\frac{1}{\sqrt{2}}(\Lambda_c\bar{\Sigma}_c^{(*)}-\Sigma_c^{(*)}\bar{\Lambda}_c)$
and $\{\Lambda_c\bar{\Sigma}_c^{(*)}\}=\frac{1}{\sqrt{2}}(\Lambda_c
\bar{\Sigma}_c^{(*)}+\Sigma_c^{(*)}\bar{\Lambda}_c)$.}\label{tab:
channel}
\end{table}

    \begin{table}[htbp]
        \begin{tabular}{cccc}\hline\hline
            Mesons&Mass(MeV)&Mesons&Mass(MeV)\\
            \hline
            $\Lambda_c^{+}$&2286.46&$\sigma$&600\\
            $\Sigma_c^{++}$&2453.97&$\pi^{\pm}$&139.57\\
            $\Sigma_c^{+}$&2452.65&$\pi^{0}$&134.98\\
            $\Sigma_c^{0}$&2453.75&$\eta$&547.86\\
            $\Sigma_c^{*++}$&2518.41&$\rho$&775.26\\
            $\Sigma_c^{*+}$&2517.4&$\omega$&782.66\\
            $\Sigma_c^{*0}$&2518.48&&\\
            \hline\hline
        \end{tabular}
        \caption{The masses of the charmed baryons and exchanged light mesons in the OBE potential, which are taken from the PDG \cite{ParticleDataGroup2020}.}\label{tab: mass meson}
    \end{table}

    \subsection{A brief discussion on the CSM}\label{subsec:csm}

We first briefly introduce the CSM before investigating the
analyticity of the OPE potentials. Aguilar, Balslev, and Combes
proposed this method in the 1970s
\cite{aguilarClassAnalyticPerturbations1971,balslevSpectralPropertiesManybody1971b},
and the corresponding conclusion is known as the ABC theorem. In
this powerful method, the resonances can be solved in the same way
as the bound states. The transformation of the radial coordinate $r$
and its conjugate momentum $k$ in the CSM is defined by:
    \begin{eqnarray}
        U(\theta)r=re^{i\theta},\qquad U(\theta)k=ke^{-i\theta}. \label{eq:rktrans}
    \end{eqnarray}
    Then the radial Schr\"odinger equation is transformed as
    \begin{eqnarray}
        &&\Big\{\frac{1}{2m}\Big[-\frac{d^2}{dr^2}+\frac{l(l+1)}{r^2}\Big]e^{-2i\theta}+V(re^{i\theta})\Big\}\psi_l^\theta(r)\nonumber\\
        &&=E(\theta)\psi_l^\theta(r). \label{eq:SECSM}
    \end{eqnarray}

\begin{figure}[htbp]
        \includegraphics[width=210pt]{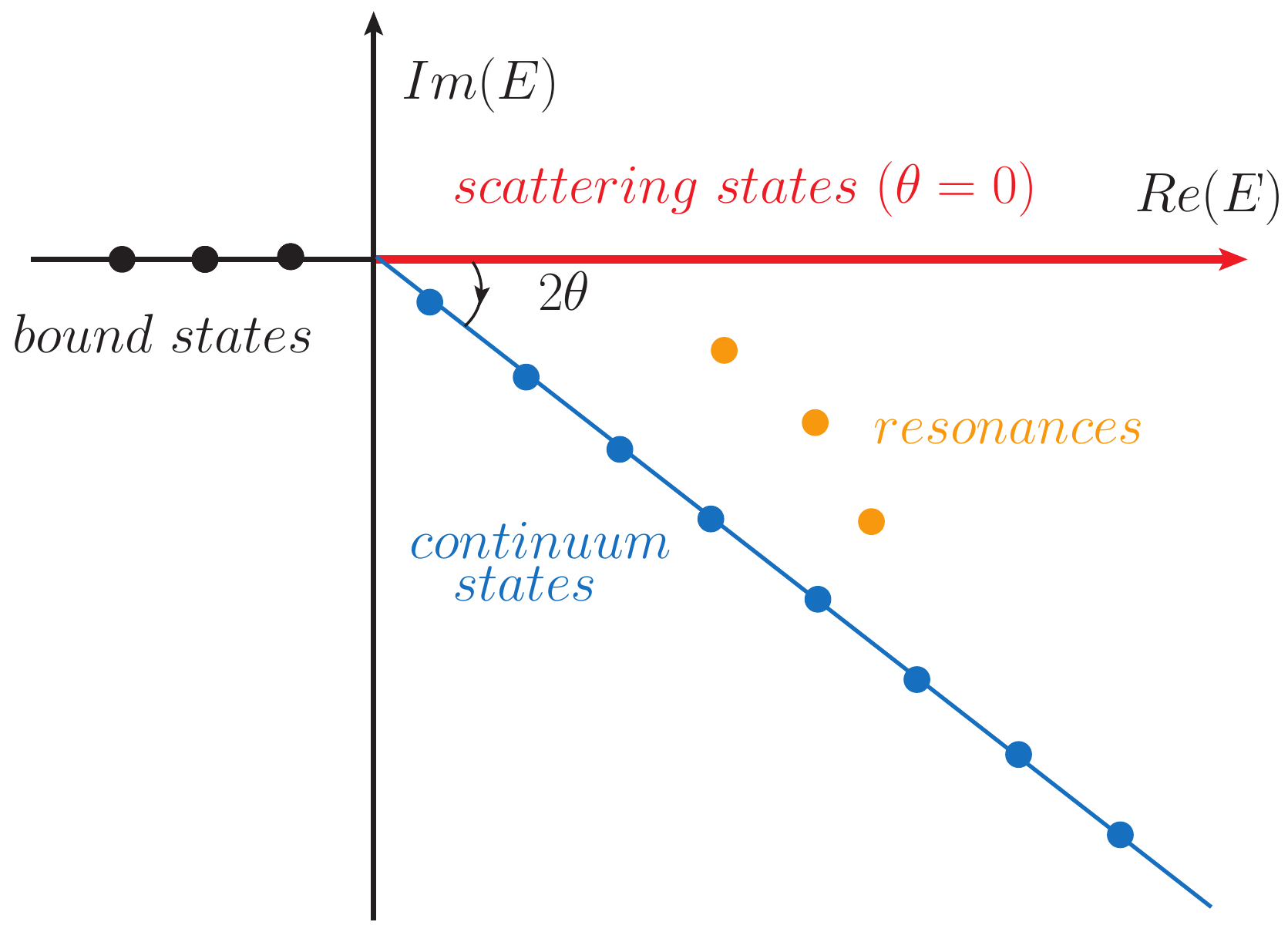}
        \caption{The eigenvalue distribution of the complex scaled Schr\"odinger equation for two-body systems. }\label{fig: CSM plot}
    \end{figure}

After the complex scaling operation, the resonance pole would cross
the branch cut into the first Riemann sheet if the rotation angle
$\theta$ is large enough, as shown in Fig. \ref{fig: CSM plot}. In
this way, the wave functions of the resonances become
square-integrable, just like the normalizable bound states. The
details can be seen in Ref.
\cite{aoyamaComplexScalingMethod2006,hoMethodComplexCoordinate1983}.

In the previous work \cite{Cheng2022b}, we adopted the Gaussian
expansion method (GEM) \cite{hiyamaGaussianExpansionMethod2003} to
solve the tetraquark system and our results agreed with the
experimental data very well. However, when dealing with resonances,
the GEM may not be applicable to some extreme situations. For
example, if there is a pole located on the second Riemann sheet
corresponding to one of the channels, we need to make a complex
scaling to move this pole to the first Riemann sheet. And when this
pole is too close to or below the threshold, the rotation angle
$\theta\gtrsim\pi/4$, which is out of the limit of the Gaussian
basis. So we adopt another function as the basis sets--- the
Laguerre functions of the form
    \begin{eqnarray}
    &&\phi_{nl}(\lambda r)=\sqrt{\frac{n!}{(2l+2+n)!}}(\lambda r)^{l+1}e^{-\lambda r/2}L_n^{2l+2}(\lambda r),\nonumber\\
    \end{eqnarray}
where the $\lambda$ is an adjustable parameter for the different
size state. The radial wave function can be expanded as
    \begin{eqnarray}
    \psi_{l}^{\theta}(r)=\sum_{n}^{N}c_{n}(\theta)\phi_{nl}(\lambda r), \label{eq:psifun}
    \end{eqnarray}
where the $c_n(\theta)$ is the rotation angle $\theta$-dependent
coefficient. These basis sets have some good characteristics: 1) We
could get all the resonances located on the second Riemann sheets
since the angle region becomes $0<\theta<\pi/2$. 2) The basis
functions are orthonormal. 3) One could evaluate all the Hamiltonian
matrix elements with simple analytical formulas in the OBE potential
case. 4) These basis sets could allow the wave function to have the
oscillating behavior of trigonometric function in a infinite range so that one could get the
partial width of the corresponding resonance with the Golden rule \cite{noroResonancePartialWidths1980,rescignoNormalizationResonanceWave1986}.
One could find the concrete application in Ref.
\cite{Wendoloski1978}.

%These basis sets contain the exponentially decaying factor
%$e^{-kr}$, which becomes $sin (k'r)$ etc after the complex scaling.
%Hence, these basis sets ensure the wave function to have the
%oscillating behavior in the infinite range so that one could get the
%partial width of the corresponding resonance with the Golden rule
%\cite{noroResonancePartialWidths1980,rescignoNormalizationResonanceWave1986}.
%One could find the concrete application in Ref.
%\cite{Wendoloski1978}.

\subsection{Analyticity of the OPE potentials for the $\Lambda_c\Sigma_c^{(*)}$ system}\label{subsec:anlyticity Hcc}

\begin{figure}[htbp]
    \subfigure[]{ \label{fig: FD Hcc}
        \includegraphics[width=180pt]{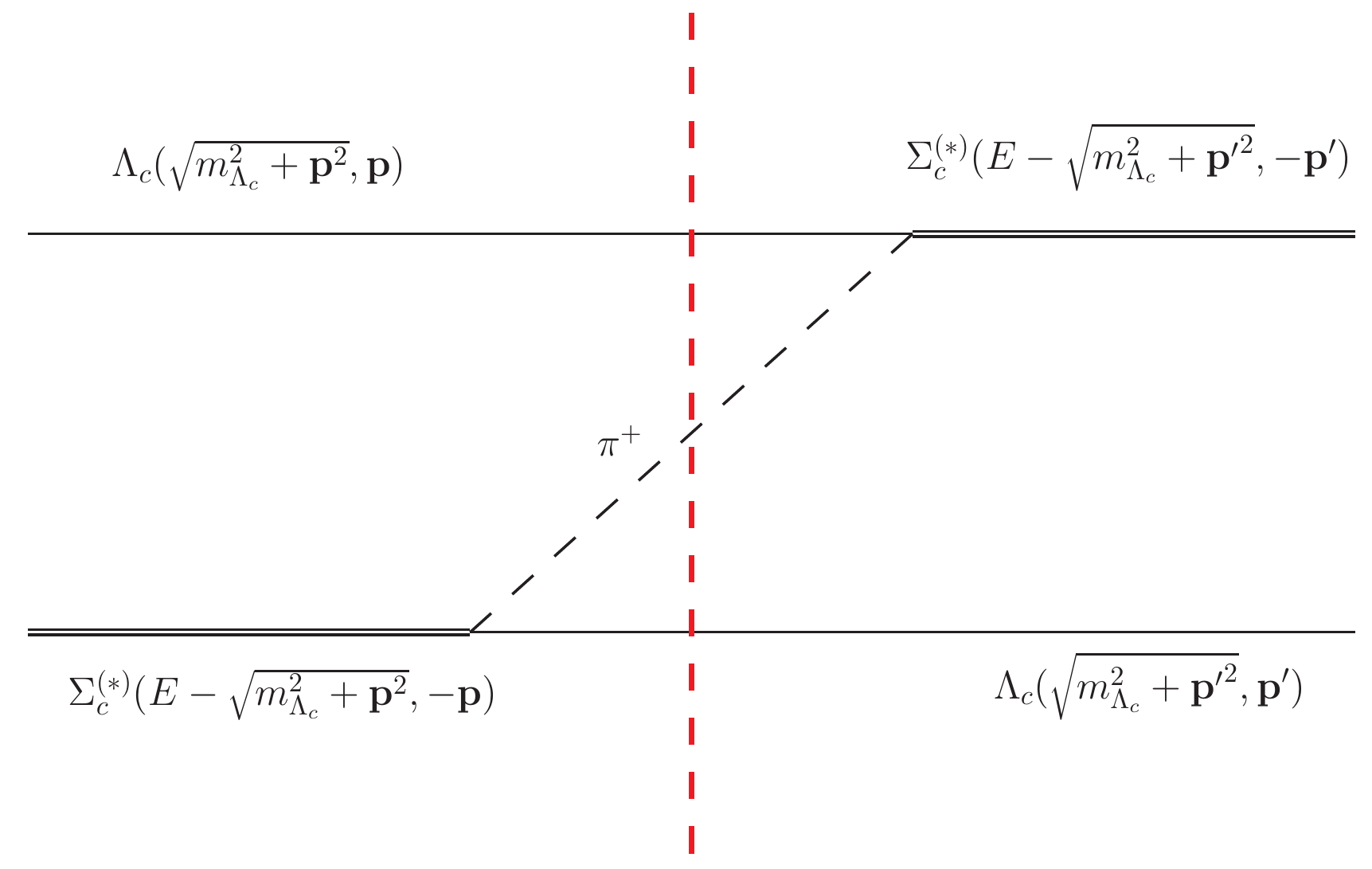}}\hspace{40pt}
    \subfigure[]{ \label{fig: FD Hcc2}
        \includegraphics[width=180pt]{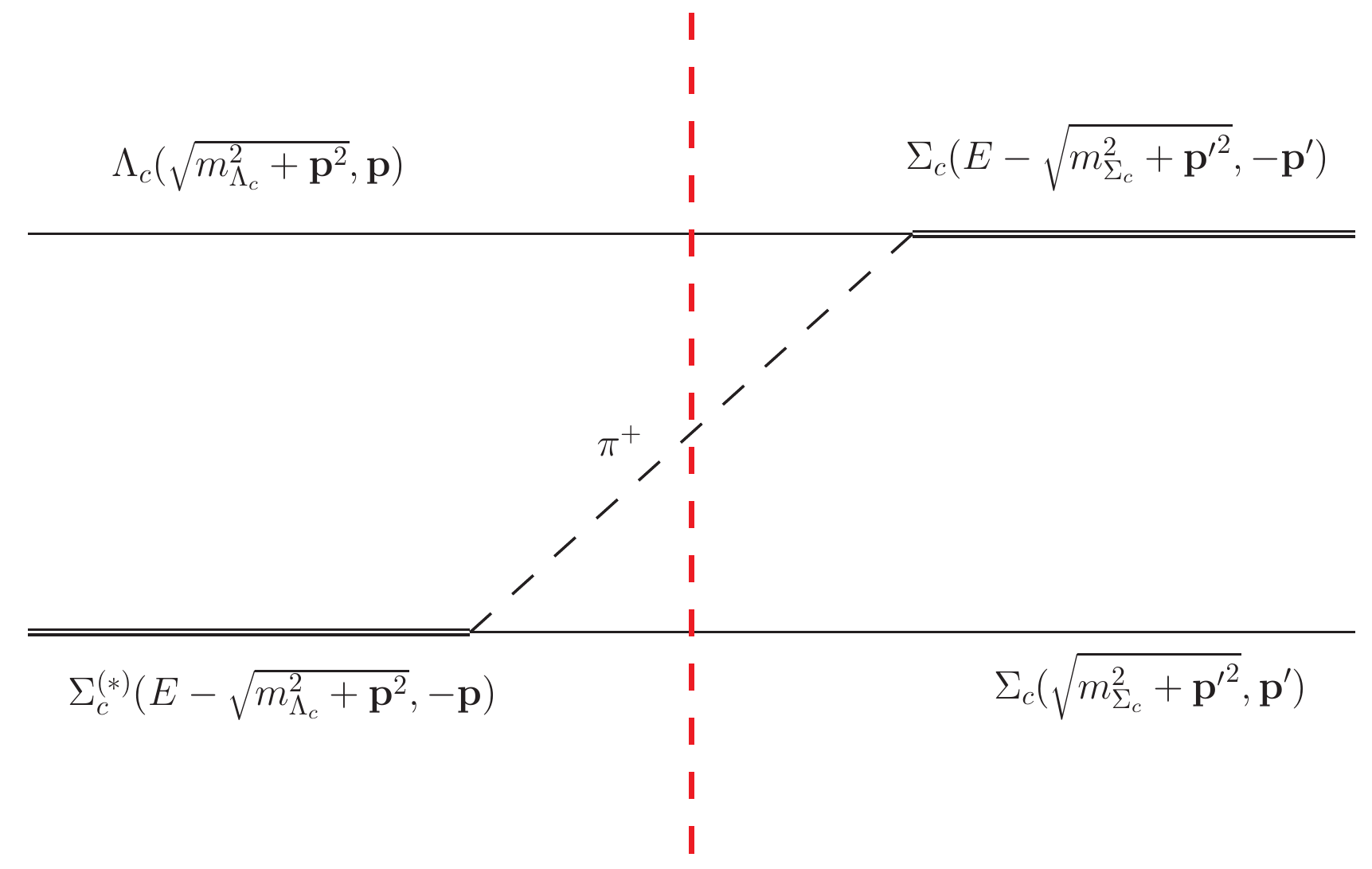}}
\caption{Three-body intermediate state diagram in the processes (a)
$\Lambda_c\Sigma_c^{(*)}\to\Sigma_c^{(*)}\Lambda_c$ and (b)
$\Lambda_c\Sigma_c^{(*)}\to\Sigma_c\Sigma_c$. We assume the total
energy is $E$, and the baryons cut by the red dashed line are on
shell.}
\end{figure}

When considering the process
$\Lambda_c\Sigma_c\to\Sigma_c\Lambda_c$, one can get potentials as
follows
    \begin{eqnarray}
        V_\pi\propto\frac{1}{2f_\pi^2}\frac{(\boldsymbol{\sigma}\cdot\boldsymbol{q})(\boldsymbol{\sigma}\cdot\boldsymbol{q})}{q^2-m_\pi^2}, \label{eq:OPE}
    \end{eqnarray}
where the $\boldsymbol{\sigma}$ is the Pauli matrix. The $q$ is the
transferred momentum of the pion, and the $q_0$ is its $0$th
component. The denominator above gives
$q^2-m_\pi^2=-(\boldsymbol{q}^2-m_{eff}^2)$, where the shorthand
$m_{eff}=\sqrt{q_0^2-m_\pi^2}$ and the $q_0\approx
m_{\Sigma_c}-m_{\Lambda_c}>m_\pi$. Obviously, the poles are located
on the real transferred momentum axis. When making a Fourier
transformation, we adopt Feynman prescription to make the contour
integral, and the OPE potential is proportional to
$1/(\boldsymbol{p}^2-m_{eff}^2-i\epsilon)$.   It is obvious that the
complex scaling operation will not change the analyticity of the OPE
potential. Compared with the $DD^*$ system, we have an additional
channel $\Lambda_c\Sigma_c^*$ that could provide a similar
potential. The processes $\Lambda_c\Sigma_c^*\to\Sigma_c^*\Lambda_c$
and $\Lambda_c\Sigma_c\to\Sigma_c^*\Lambda_c$ could contribute an
imaginary part too. To discuss this type of process, we make a
careful discussion on the $q_0$ herein. Since
$m_{\Sigma_c}-m_{\Lambda_c}$ and $m_\pi$ are comparable, the
effective mass $m_{eff}\approx \sqrt{2m_\pi
(m_{\Sigma_c}-m_{\Lambda_c}-m_{\pi})}$ is small. Therefore, the
small bound energy could also affect the $m_{eff}$. To deal with the
$q_0$ in the $\Lambda_c\Sigma_c^{(*)}\to\Sigma_c^{(*)}\Lambda_c$
process, we denote the total energy as $E$ and assume the
$\Lambda_c$ to be on shell. Then the
$q_0=E-\sqrt{m_{\Lambda_c}^2+\boldsymbol{p}^2}-\sqrt{m_{\Lambda_c}^2+\boldsymbol{p'}^2}$,
as illustrated in Fig. \ref{fig: FD Hcc}.

We will neglect the kinetic energy terms
$\boldsymbol{p}^2/2m_{\Lambda_c}$ and
$\boldsymbol{p}'^2/2m_{\Lambda_c}$ of the charmed baryons due to the
heavy quark approximation. Then we make an energy shift $E\to
E+2m_{\Lambda_c}$, and the $q_0$ gives
$q_0=E+m_{\Sigma_c}-m_{\Lambda_c}$. For the process containing the
$\Sigma_c\Sigma_c$ channel, we take $q_0=0$ in the diagonal process
$\Sigma_c\Sigma_c\to\Sigma_c\Sigma_c$ and $q_0=E$ in the
non-diagonal process $\Lambda_c\Sigma_c^{(*)}\to\Sigma_c\Sigma_c$.
We illustrate the latter one in Fig. \ref{fig: FD Hcc2}. In this
work, the latter one could also provide an imaginary part when the
energy is around the threshold of the $\Sigma_c\Sigma_c$. In fact,
we also use these assumptions in the process with other propagators.
We list $q_0$ values for different cases in Table \ref{tab: q0}. To
distinguish the special $q_0$ in the cross diagrams
$\Lambda_c\Sigma_c^{(*)}\to\Sigma_c^{(*)}\Lambda_c$ from the $q_0$
in the direct diagrams
$\Lambda_c\Sigma_c^{(*)}\to\Lambda_c\Sigma_c^{(*)}$, we use the
shorthand "$q_0^C$" for the former cases when we give concrete
expressions of the potentials.

\begin{table}[htbp]
        \begin{tabular}{ccc}\hline\hline
            Process&$\Lambda_c\Sigma_c^{(*)}\to\Sigma_c^{(*)}\Lambda_c$&$\Lambda_c\Sigma_c^{(*)}\to\Sigma_c\Sigma_c$\\
            $q_0$&$E+m_{\Sigma_c}-m_{\Lambda_c}$&$E$\\
            \hline
             Process&$\Lambda_c\bar{\Sigma}_c^{(*)}\to\Sigma_c^{(*)}\bar{\Lambda}_c$&$\Lambda_c\bar{\Sigma}_c^{(*)}\to\Sigma_c\bar{\Sigma}_c$\\
            $q_0$&$E+m_{\Sigma_c}-m_{\Lambda_c}$&$E$\\
            \hline\hline
        \end{tabular}
\caption{The $q_0$ is the 0th component of the transferred momentum.
The $E$ is the total energy relative to the threshold of the
$\Lambda_c\Sigma_c$. The $q_0=E+m_{\Sigma_c}-m_{\Lambda_c}$ is from
the cross diagram
$\Lambda_c\Sigma_c^{(*)}\to\Sigma_c^{(*)}\Lambda_c$. And the $q_0$
from the direct diagram
$\Lambda_c\Sigma_c^{(*)}\to\Lambda_c\Sigma_c^{(*)}$ is equal to 0.
The cases not listed all give $q_0=0$.}\label{tab: q0}
\end{table}

\section{Lagrangians and Potentials}\label{sec: potentials}

The effective Lagrangians are built under the heavy quark symmetry
and SU(3)-flavor symmetry. The concrete expressions of the OBE
Lagrangians read as
    \begin{eqnarray}
        \mathcal{L}_{B} & = & \mathcal{L}_{B_{\bar{3}}}+\mathcal{L}_{S}+\mathcal{L}_{int}, \\
        \mathcal{L}_{B_{\bar{3}}} & = & \frac{1}{2}\langle\bar{B}_{\bar{3}}(iv\cdot D)B_{\bar{3}}\rangle+i\beta_B\langle\bar{B}_{\bar{3}}v^\mu(\Gamma_\mu-V_\mu)B_{\bar{3}} \rangle+l_B\langle\bar{B}_{\bar{3}}\sigma B_{\bar{3}}\rangle,\label{lag-hb-B3}\\
        \mathcal{L}_{S} & =& -\langle\bar{S}^\alpha(iv\cdot D-\Delta_B)S_\alpha\rangle+\frac{3}{2}g_{1}(iv_{k})\epsilon^{\mu\nu\lambda\kappa}\langle \bar{S}_{\mu}\mathbb{A}_{\nu}S_{\lambda}\rangle+i\beta_S\langle\bar{S}_{\mu}v_\alpha(\Gamma^\alpha-V^\alpha)S_{\mu} \rangle\nonumber\\
        &+&\lambda_S\langle\bar{S}_{\mu}F^{\mu\nu}S_{\nu}\rangle+l_S\langle\bar{S}_{\mu}\sigma S_{\mu}\rangle,\\
        \mathcal{L}_{int} & = & g_{4}\langle\bar{S}^{\mu}\mathbb{A}_{\mu}B_{\bar{3}}\rangle+i\lambda_I \epsilon^{\mu\nu\lambda\kappa}v_\mu\langle\bar{S}_{\nu}F^{\lambda\kappa}B_{\bar{3}}\rangle+h.c..
    \end{eqnarray}
The $S^\mu$ and$B_{\bar{3}}$ are the heavy sextet and anti-triplet
baryon superfield, defined as
    \begin{eqnarray}
    S_{\mu} &=& B_{\mu}^{*}-\frac{1}{\sqrt{3}}(\gamma_{\mu}+v_{\mu})\gamma^{5}B_{6}.
    \end{eqnarray}
These heavy baryon fields are {\small
    \begin{align}\label{field-HB}
        & B_{\bar{3}} =
        \begin{pmatrix}
            0 & \Lambda_{Q} & \Xi_{Q}^{+1/2} \\
            -\Lambda_{Q} & 0 & \Xi_{Q}^{-1/2} \\
            \Xi_{Q}^{+1/2} & \Xi_{Q}^{-1/2} & 0
        \end{pmatrix},
        B_{6} =
        \begin{pmatrix}
            \Sigma_{Q}^{+1} & \frac{1}{\sqrt{2}}\Sigma_{Q}^{0} & \frac{1}{\sqrt{2}}\Xi_{Q}^{'+1/2} \\
            \frac{1}{\sqrt{2}}\Sigma_{Q}^{0} & \Sigma_{Q}^{-1} & \frac{1}{\sqrt{2}}\Xi_{Q}^{'-1/2} \\
            \frac{1}{\sqrt{2}}\Xi_{Q}^{'+1/2} & \frac{1}{\sqrt{2}}\Xi_{Q}^{'-1/2} & \Omega_{Q}
        \end{pmatrix},\nonumber\\
        & B_{6}^{*} =
        \begin{pmatrix}
            \Sigma_{Q}^{*+1} & \frac{1}{\sqrt{2}}\Sigma_{Q}^{*0} & \frac{1}{\sqrt{2}}\Xi_{Q}^{*'+1/2} \\
            \frac{1}{\sqrt{2}}\Sigma_{Q}^{*0} & \Sigma_{Q}^{*-1} & \frac{1}{\sqrt{2}}\Xi_{Q}^{*'-1/2} \\
            \frac{1}{\sqrt{2}}\Xi_{Q}^{*'+1/2} & \frac{1}{\sqrt{2}}\Xi_{Q}^{*'-1/2} & \Omega_{Q}^{*}
        \end{pmatrix}.
\end{align}}
The light meson part are given below
    \begin{eqnarray}
    \mathbb{A} & = &\frac{i}{2} [\xi^\dagger(\partial_\mu\xi)+(\partial_\mu\xi)\xi^\dagger],\quad \Gamma_\mu=\frac{1}{2} [\xi^\dagger(\partial_\mu\xi)-(\partial_\mu\xi)\xi^\dagger],\quad F_{\mu\nu}=\partial_\mu V_\nu-\partial_\nu V_\mu+[V_\mu,V_\nu],\nonumber \\
    \xi & = &\text{exp}[\frac{i\mathcal{M}}{f_\pi}],\nonumber \\
    \xi & = & \begin{pmatrix}
        \frac{\pi^0}{\sqrt{2}}+\frac{\eta}{\sqrt{6}} & \pi^+ & K^+ \\
        \pi^- & \frac{\pi^0}{\sqrt{2}}+\frac{\eta}{\sqrt{6}} & K^0 \\
        K^- & \bar{K}^0 & -\frac{2}{\sqrt{6}}\eta
    \end{pmatrix}, \quad
V^\mu=i\frac{g_V}{\sqrt{2}}\begin{pmatrix}
    \frac{\rho^0}{\sqrt{2}}+\frac{\omega}{\sqrt{2}} & \rho^+ & K^{*+} \\
    \rho^- & -\frac{\rho^0}{\sqrt{2}}+\frac{\eta}{\sqrt{2}} & K^{*0} \\
    K^{*-} & \bar{K}^{*0} & \phi
\end{pmatrix}.
\end{eqnarray}

For the pion decay constant, we use $f_\pi=132$ MeV. As for the
other coupling constants, we adopt the values in Ref.
\cite{Liu2012}:
    \begin{eqnarray}
    g_2 & = &-0.598,\quad g_4=0.999, \quad g_1=\frac{8}{3}g_4,\nonumber \\
    g_3 & = &\sqrt{\frac{2}{3}}g_4,\quad g_5=-\sqrt{2}g_4,\quad g_A=1.25,\nonumber \\
    l_B & = &-3.1,\quad l_S=-2l_B, \quad \beta_B g_V=-6.0,\nonumber \\
    \beta_S g_V & = &-2\beta_S g_V,\quad \lambda_S g_V=19.2 \text{GeV}^{-1}, \quad \lambda_I g_V=-\lambda_S g_V/\sqrt{8}.\nonumber \\
\end{eqnarray}

To get the effective potentials, we add a monopole form factor at
each vertex
\begin{eqnarray}
    F_i(q)=\frac{\Lambda_i^2-m_i^2}{\Lambda_i^2-q^2},\label{eq:form factor}
\end{eqnarray}
where the $i$ stands for one of the propagators ($\pi$, $\eta$,
$\sigma$, $\rho$ and $\omega$), the $q^2=q_0^2-\boldsymbol{q}^2$,
and the $\Lambda_i$ and $m_i$ are the cutoff parameter and the mass
of the corresponding propagator. After the Fourier transformation
\begin{eqnarray}
    V_i(r)=\frac{1}{(2\pi)^3}\int d\boldsymbol{q}^3e^{-i\boldsymbol{q}\cdot\boldsymbol{r}}V(\boldsymbol{q})F_i^2(q),\label{eq:FT}
\end{eqnarray}
we can get the coordinate space potentials {\small
    \begin{eqnarray}
    V^{\Lambda_c\Sigma_c\to\Lambda_c\Sigma_c}&=&-\frac{g_2^2}{f_\pi^2}\left[S(\boldsymbol{\sigma}_1,\boldsymbol{\sigma}_2)Y_3(\Lambda_\pi,q_0^C,m_\pi,r)+T(\boldsymbol{\sigma}_1,\boldsymbol{\sigma}_2)H_3(\Lambda_\pi,q_0^C,m_\pi,r)\right]\boldsymbol{\epsilon}_3^\dagger\cdot\boldsymbol{\epsilon}_2+2l_B l_S Y_0(\Lambda_\sigma,q_0,m_\sigma,r)\nonumber\\
    &-&\frac{1}{3}(\lambda_I g_V)^2\left[2S(\boldsymbol{\sigma}_1,\boldsymbol{\sigma}_2)Y_3(\Lambda_\rho,q_0^C,m_\rho,r)-T(\boldsymbol{\sigma}_1,\boldsymbol{\sigma}_2)H_3(\Lambda_\rho,q_0^C,m_\rho,r)\right]\boldsymbol{\epsilon}_3^\dagger\cdot\boldsymbol{\epsilon}_2\nonumber\\
    &-&\frac{1}{2}(\beta_B \beta_S g_V^2)Y_0(\Lambda_\omega,q_0,m_\omega,r),\\
    V^{\Lambda_c\Sigma_c^*\to\Lambda_c\Sigma_c^*}&=&-\frac{g_4^2}{f_\pi^2}\left[S(\boldsymbol{S}_{t3}^\dagger,\boldsymbol{S}_{t2})Y_3(\Lambda_\pi,q_0^C,m_\pi,r)+T(\boldsymbol{S}_{t3}^\dagger,\boldsymbol{S}_{t2})H_3(\Lambda_\pi,q_0^C,m_\pi,r)\right]\boldsymbol{\epsilon}_3^\dagger\cdot\boldsymbol{\epsilon}_2+2l_B l_S Y_0(\Lambda_\sigma,q_0,m_\sigma,r)\nonumber\\
    &+&2(\lambda_I g_V)^2\left[2S(\boldsymbol{S}_{t3}^\dagger,\boldsymbol{S}_{t2})Y_3(\Lambda_\rho,q_0^C,m_\rho,r)-T(\boldsymbol{S}_{t3}^\dagger,\boldsymbol{S}_{t2})H_3(\Lambda_\rho,q_0^C,m_\rho,r)\right]\boldsymbol{\epsilon}_3^\dagger\cdot\boldsymbol{\epsilon}_2\nonumber\\
&+&\frac{1}{2}(\beta_B \beta_S g_V^2)Y_0(\Lambda_\omega,q_0,m_\omega,r),\\
    V^{\Sigma_c\Sigma_c\to\Sigma_c\Sigma_c}&=&\frac{g_1^2}{2f_\pi^2}\left[S(\boldsymbol{\sigma}_1,\boldsymbol{\sigma}_2)Y_3(\Lambda_\pi,q_0,m_\pi,r)+T(\boldsymbol{\sigma}_1,\boldsymbol{\sigma}_2)H_3(\Lambda_\pi,q_0,m_\pi,r)\right]\boldsymbol{I}_1\cdot\boldsymbol{I}_2\nonumber\\
    &+&\frac{1}{3}\frac{g_1^2}{2f_\pi^2}\left[S(\boldsymbol{\sigma}_1,\boldsymbol{\sigma}_2)Y_3(\Lambda_\eta,q_0,m_\eta,r)+T(\boldsymbol{\sigma}_1,\boldsymbol{\sigma}_2)H_3(\Lambda_\eta,q_0,m_\eta,r)\right]-l_S^2 Y_0(\Lambda_\sigma,q_0,m_\sigma,r)\nonumber\\
    &+&\frac{1}{2}(g_S g_V)^2Y_0(\Lambda_\rho,q_0,m_\rho,r)\boldsymbol{I}_1\cdot\boldsymbol{I}_2-\frac{1}{3}(\lambda_S g_V)^2\left[2S(\boldsymbol{\sigma}_1,\boldsymbol{\sigma}_2)Y_3(\Lambda_\rho,q_0,m_\rho,r)-T(\boldsymbol{\sigma}_1,\boldsymbol{\sigma}_2)H_3(\Lambda_\rho,q_0,m_\rho,r)\right]\boldsymbol{I}_1\cdot\boldsymbol{I}_2\nonumber\\
    &+&\frac{1}{2}(\beta_S g_V)^2Y_0(\Lambda_\omega,q_0,m_\omega,r)-\frac{1}{3}(\lambda_S g_V)^2\left[2S(\boldsymbol{\sigma}_1,\boldsymbol{\sigma}_2)Y_3(\Lambda_\omega,q_0,m_\omega,r)-T(\boldsymbol{\sigma}_1,\boldsymbol{\sigma}_2)H_3(\Lambda_\omega,q_0,m_\omega,r)\right],\\
    V^{\Lambda_c\Sigma_c\to\Lambda_c\Sigma_c^*}&=&-\frac{g_2 g_4}{f_\pi^2}\left[S(\boldsymbol{S}_{t3}^\dagger,\boldsymbol{\sigma}_2)Y_3(\Lambda_\pi,q_0^C,m_\pi,r)+T(\boldsymbol{S}_{t3}^\dagger,\boldsymbol{\sigma}_2)H_3(\Lambda_\pi,q_0^C,m_\pi,r)\right]\boldsymbol{\epsilon}_3^\dagger\cdot\boldsymbol{\epsilon}_2\nonumber\\
    &+&\frac{2}{\sqrt{3}}(\lambda_I g_V)^2\left[2S(\boldsymbol{S}_{t3}^\dagger,\boldsymbol{\sigma}_2)Y_3(\Lambda_\rho,q_0^C,m_\rho,r)-T(\boldsymbol{S}_{t3}^\dagger,\boldsymbol{\sigma}_2)H_3(\Lambda_\rho,q_0^C,m_\rho,r)\right]\boldsymbol{\epsilon}_3^\dagger\cdot\boldsymbol{\epsilon}_2,\\
    V^{\Lambda_c\Sigma_c\to\Sigma_c\Sigma_c}&=&\frac{g_1 g_2}{\sqrt{2}f_\pi^2}\left[S(\boldsymbol{\sigma}_{1},\boldsymbol{\sigma}_2)Y_3(\Lambda_\pi,q_0,m_\pi,r)+T(\boldsymbol{\sigma}_{1},\boldsymbol{\sigma}_2)H_3(\Lambda_\pi,q_0,m_\pi,r)\right]\boldsymbol{\epsilon}_3^\dagger\cdot\boldsymbol{I}_2\nonumber\\
    &+&\frac{2}{3\sqrt{6}}(\lambda_I\lambda_S g_V^2)\left[2S(\boldsymbol{\sigma}_{1},\boldsymbol{\sigma}_2)Y_3(\Lambda_\rho,q_0,m_\rho,r)-T(\boldsymbol{\sigma}_{1}^\dagger,\boldsymbol{\sigma}_2)H_3(\Lambda_\rho,q_0,m_\rho,r)\right]\boldsymbol{\epsilon}_3^\dagger\cdot\boldsymbol{I}_2,\\
    V^{\Lambda_c\Sigma_c^*\to\Sigma_c\Sigma_c}&=&\frac{g_2 g_3}{\sqrt{2}f_\pi^2}\left[S(\boldsymbol{\sigma}_{1},\boldsymbol{S}_{t2})Y_3(\Lambda_\pi,q_0,m_\pi,r)+T(\boldsymbol{\sigma}_{1},\boldsymbol{S}_{t2})H_3(\Lambda_\pi,q_0,m_\pi,r)\right]\boldsymbol{\epsilon}_3^\dagger\cdot\boldsymbol{I}_2\nonumber\\
    &-&\frac{1}{3\sqrt{2}}(\lambda_I\lambda_S g_V^2)\left[2S(\boldsymbol{\sigma}_{1},\boldsymbol{S}_{t2})Y_3(\Lambda_\rho,q_0,m_\rho,r)-T(\boldsymbol{\sigma}_{1},\boldsymbol{S}_{t2})H_3(\Lambda_\rho,q_0,m_\rho,r)\right]\boldsymbol{\epsilon}_3^\dagger\cdot\boldsymbol{I}_2.\\
    \end{eqnarray} }
We added a factor -1 for the cross-diagram potentials, which contain
the $q_0^C$. The factor is from the fermions position exchange, and
equal to $(-1)^{s-s_1-s_2+l+i-i_1-i_2+1}$, where the
$s,s_1,s_2,l,i,i_1,i_2$ are spin, orbit and isospin numbers. The
$\boldsymbol{\epsilon}$ and $\boldsymbol{I}$ are the isospin
polarization vector and isospin operator respectively, and the
isospin-dependent matrix elements are given in Table \ref{tab:
isospin matrix element}. The $\boldsymbol{S}_t$ and
$\boldsymbol{\sigma}$ are the spin transition operator and Pauli
operator respectively. The spin-dependent operators have
$S(\mathbf{a},\mathbf{b})=\mathbf{a}\cdot\mathbf{b}$ and
$T(\mathbf{a},\mathbf{b})=3(\mathbf{a}\cdot\mathbf{r})(\mathbf{b}\cdot\mathbf{r})/r^2-\mathbf{a}\cdot\mathbf{b}$,
whose matrix elements are given in Table \ref{tab: spin matrix
element}. The $Y_3$, $H_3$ functions and relevant $Y$, $H$ functions
are defined as
    \begin{eqnarray}
        &&Y(x)=\frac{e^{-x}}{x},\quad H(x)=(1+\frac{3}{x}+\frac{3}{x^2})Y(x),\nonumber\\
        &&Y_0(\Lambda,q_0,m,r)=\frac{u}{4\pi}[Y(ur)-\frac{\chi}{u}Y(\chi r)-\frac{\beta^2}{2\chi u}e^{-\chi r}],\nonumber\\
        &&Y_3(\Lambda,q_0,m,r)=\frac{u^3}{12\pi}[Y(ur)-\frac{\chi}{u}Y(\chi r)-\frac{\beta^2\chi}{2u^3}e^{-\chi r}],\nonumber\\
        &&H_3(\Lambda,q_0,m,r)=\frac{u^3}{12\pi}[H(ur)-(\frac{\chi}{u})^3 H(\chi r)\nonumber\\
        &&\qquad\qquad\qquad-\frac{\beta^2}{2\chi u}\frac{\chi^2}{u^2}Y(\chi r)-\frac{\beta^2}{2\chi u}\frac{\chi^2}{u^2}e^{-\chi r}],\label{eq:YZH}
    \end{eqnarray}
    where the
    \begin{eqnarray}
        &&u=\text{Sign}\big[\text{Re}\big(e^{i\theta}\sqrt{m^2-q_0^2}\big)\big]\sqrt{m^2-q_0^2}, \nonumber\\
        &&\beta=\sqrt{\Lambda^2-m^2},\quad \chi=\sqrt{\Lambda^2-q_0^2}.
    \end{eqnarray}

    \begin{table}[htbp]
        \caption{The spin-dependent matrix elements. }\label{tab: spin matrix element}
        \begin{tabular}{ccccccccc}\hline\hline
            $\Delta$&$S(\boldsymbol{\sigma}_1^\dagger,\boldsymbol{\sigma}_2)$&$T(\boldsymbol{\sigma}_1^\dagger,\boldsymbol{\sigma}_2)$&$S(\boldsymbol{S}_{t3}^\dagger,\boldsymbol{S}_{t2})$&$T(\boldsymbol{S}_{t3}^\dagger,\boldsymbol{S}_{t2})$&$S(\boldsymbol{S}_{t3}^\dagger,\boldsymbol{\sigma}_{2})$&$T(\boldsymbol{S}_{t3}^\dagger,\boldsymbol{\sigma}_{2})$&$S(\boldsymbol{\sigma}_{1},\boldsymbol{S}_{t2})$&$T(\boldsymbol{\sigma}_{1},\boldsymbol{S}_{t2})$\\
            $\langle ^3S_1|\Delta|^3S_1\rangle$&1&0&$\frac{1}{3}$&$0$&$-2\sqrt{\frac{2}{3}}$&$0$&$2\sqrt{\frac{2}{3}}$&0\\
            $\langle ^3D_1|\Delta|^3S_1\rangle$&0&$2\sqrt{2}$&$0$&$-\frac{5}{3\sqrt{2}}$&0&$\frac{1}{\sqrt{3}}$&0&$-\frac{1}{\sqrt{3}}$\\
            $\langle ^3S_1|\Delta|^3D_1\rangle$&0&$2\sqrt{2}$&$0$&$-\frac{5}{3\sqrt{2}}$&0&$\frac{1}{\sqrt{3}}$&0&$-\frac{1}{\sqrt{3}}$\\
            $\langle ^3D_1|\Delta|^3D_1\rangle$&1&-2&$\frac{1}{3}$&$\frac{5}{6}$&$-2\sqrt{\frac{2}{3}}$&$-\frac{1}{\sqrt{6}}$&$2\sqrt{\frac{2}{3}}$&$\frac{1}{\sqrt{6}}$\\
            \hline
            $\Delta$&$S(\boldsymbol{\sigma}_1^\dagger,\boldsymbol{\sigma}_2)$&&&&&&&\\
            $\langle ^1S_0|\Delta|^1S_0\rangle$&-3&&&&&&&\\
            \hline\hline
        \end{tabular}
    \end{table}

    \begin{table}[htbp]
    \caption{The isospin-dependent ($I$=1) matrix elements of the operators $\boldsymbol{\epsilon}_3^\dagger\cdot\boldsymbol{\epsilon}_2$, $\boldsymbol{\epsilon}_3^\dagger\cdot\boldsymbol{I}_2$ and $\boldsymbol{I}_1\cdot\boldsymbol{I}_2$.  }\label{tab: isospin matrix element}
    \begin{tabular}{cccc}\hline\hline
        $\Delta$&$\boldsymbol{\epsilon}_3^\dagger\cdot\boldsymbol{\epsilon}_2$&$\boldsymbol{\epsilon}_3^\dagger\cdot\boldsymbol{I}_2$&$\boldsymbol{I}_1\cdot\boldsymbol{I}_2$\\
        \hline
        $\langle I=1|\Delta|I=1\rangle$&$1$&$-\sqrt{2}$&$-1$\\
        \hline\hline
    \end{tabular}
\end{table}

\section{Numerical results}\label{sec: results}
\subsection{The OPE potential results for the double-charm hexaquark system}\label{sec:OPE result}

We first introduce the OPE potentials to study the double-charm
hexaquark system. For the $1(0^+)$ system, we do not get a bound or
resonant state with a reasonable cutoff $\Lambda_\pi$. However, we
find a quasibound state in the $1(1^+)$ case, and the results are
shown in row "Adopt" of Table \ref{tab: cc OPE}. One could find this
pole has an imaginary part corresponding to $-i\Gamma/2$. We give a
brief explanation of the measure
$\langle\tilde{\psi}_i|\psi_i\rangle$ herein. The
$\langle\tilde{\psi}_i|\psi_i\rangle=e^{i\theta}\int_{0}^{\infty}\{\psi_i(re^{i\theta})\}^2dr$
is the amplitude corresponding to the $i$-th channel. This measure
is similar to the definition of the probabilities of bound states.
However, one could find that its value could be complex in Table
\ref{tab: cc OPE} and could not be regarded as probabilities. This
behavior is from the normalization of the resonance wave function
\cite{More1973,rescignoNormalizationResonanceWave1986}. However, the
quasibound state herein is special. When the bound energy $B.E.\le
-(m_{\Sigma_c}-m_{\Lambda_c}-m_\pi)$, the quasibound state turns
into a bound state, and the width from the three-body decay effect
vanishes. Then the $\langle\tilde{\psi}_i|\psi_i\rangle$ turns into
the probability of the $i$-th channel. Hence this measure could
partly reflect the constituents of this special quasibound state.
Furthermore, the $\sqrt{\langle r^2\rangle}$ is the
root-mean-square-radius. Its real part is interpreted as an
expectation value, and the imaginary part corresponds to a measure
of the uncertainty in observation
\cite{hommaMatrixElementsPhysical1997}.

To make a comparison, we also give the result under the
instantaneous approximation in row "$q_0=0$" of Table \ref{tab: cc
OPE}. In this case, the imaginary part of the OPE potential from the
three-body decay effect disappears, and the pole becomes a bound
state. Considering the cases in row "Adopt" with $\Lambda_\pi=1.0$
and row "$q_0=0$" with $\Lambda_\pi=1.05$, we find their values are
close to each other, including the energy, $\sqrt{\langle
r^2\rangle}$ and $\langle\tilde{\psi}_i|\psi_i\rangle$. In fact, we
find this conclusion could extend to other similar systems. In other
words, the influence of the three-body decay is small. Therefore, we
will take the $\sqrt{\langle r^2\rangle}$ and
$\langle\tilde{\psi}_i|\psi_i\rangle$ as the reference measures when
analyzing the size and constituents of the pole state.

\begin{table}[!htb]
        \centering
        \scalebox{1.2}{
            \begin{tabular}{c|c|ccc}  \hline
                 &  $\Lambda_{\pi}$ (GeV)   & $1.0$                 &1.05 & $1.1$  \\
                \hline
            \multirow{4}{*}{Adopt} &    Energy (MeV)            & $-5.61-0.43i$         & $-13.90-0.29i$ & $-25.57-0.04i$  \\
            &   $\sqrt{\langle r^2\rangle}$ (fm)                   & $1.4-0.1i$         &  $0.9$ & $0.7$    \\
            &   $\langle\tilde{\psi}_i|\psi_i\rangle\times 100$               & $(70.7-1.4i/1.7/23.9+1.3i$  & $(59.0-0.4i/1.2/34.1+0.4$ & ($50.8$/$0.8$/$40.8$  \\
            &                           & $/0.9/2.5+0.1i/0.3)$   & $1.2/4.2/0.3$) & /$1.4$/$6.0$/$0.3$) \\
                \hline
            \multirow{4}{*}{$q_0=0$} &    Energy (MeV)            & $-0.54$         & $-5.64$ & $-16.51$  \\
            &   $\sqrt{\langle r^2\rangle}$ (fm)                   & $4.1$         &  $1.3$ & $0.8$    \\
            &   $\langle\tilde{\psi}_i|\psi_i\rangle\times 100$               & $(91.4/1.0/6.4$  & $(71.7/1.2/22.8$ & ($56.8$/$0.8$/$35.4$  \\
            &   & $/0.3/0.7/0.2)$   & $(0.8/3.1/0.4$ & /$1.2$/$5.5$/$0.3$) \\
            \hline
            \end{tabular}
        }
        \centering
\caption{\small  Solutions for the double-charm hexaquark with
$I(J^{P})=1(1^+)$ in the OPE potential case with the
$\theta=20^{\circ}$. The energies are given relative to the
threshold of $\Lambda_c\Sigma_c$. The $\sqrt{\langle
r^2\rangle}=[e^{3i\theta}\int_{0}^{\infty}\{\psi(re^{i\theta})\}^2
r^2dr]^{1/2}$ is the root-mean-square (RMS) radius. The
$\langle\tilde{\psi}_i|\psi_i\rangle=e^{i\theta}\int_{0}^{\infty}\{\psi_i(re^{i\theta})\}^2dr$
is the amplitude corresponding to the $i$-th channel of
$\Lambda_c\Sigma_c(^3S_1,^3D_1)$, $\Lambda_c
\Sigma_c^*(^3S_1,^3D_1)$, $\Sigma_c \Sigma_c(^3S_1,^3D_1)$. The data
of the row "Adopt" are the results we actually adopt, and the $q_0$
herein is from Table \ref{tab: q0}. The data of row "$q_0=0$" are
from the instantaneous approximation. }\label{tab: cc OPE}
\end{table}

In our framework, the width of a quasibound state should not be
larger than the value of $\Sigma_c$ and will decrease as the binding
energy becomes deeper. In fact, so it does. Taking the case
$\Lambda_\pi=1.0$ GeV as an example, we get a quasibound state whose
binding energy and width are -5.61 MeV and 0.86 MeV respectively.
And its dominant constituents are $S$-wave $\Lambda_c\Sigma_c$ and
$\Lambda_c\Sigma_c^*$. In fact, the main channel
$\Lambda_{c}\Sigma_c(^3S_1)$ contributes a repulsive potential, and
the $\Lambda_{c}\Sigma_c^*(^3S_1)$ contributes an attractive
potential with an imaginary part. Considering only the former
channel, one could not get a bound state. However, one could get a
quasibound state when adding the attractive potential of channel
$\Lambda_c\Sigma_c^*$. Obviously, the coupled-channel effects play
an important role in this system. To gain a deeper insight, we would
add the other medium- and short-range potentials in the following
parts.

\subsection{The OBE potential results for the double-charm hexaquark system}

In this part, we further employ the OBE potential to include the
short- and medium-range contribution. Before making a calculation,
we need to choose the cutoff for the different vertex first. The
same cutoff is usually used for all the OBE potentials, like
$\Lambda_\pi=\Lambda_\eta=\Lambda_\sigma=\Lambda_\rho=\Lambda_\omega$
(common cutoff). Another possible choice could be found in Refs.
\cite{Cheng2005,liuBoundStatesRevisited2012a}. The authors adopted
$\Lambda_i=m_i+\alpha\Lambda_{\text{QCD}}$ (scaled cutoff), where
the $i$ is corresponds to a propagator ($\pi$, $\eta$, $\sigma$,
$\rho$, $\omega$), $\Lambda_{\text{QCD}}=220$ MeV is the scale of
QCD, and $\alpha$ is a dimensionless parameter. However, these two
choices could both be invalid in this work.

As shown in Table \ref{tab: cc OPE}, the cutoff in the OPE potential
case is close to 1 GeV. And when we add the other potentials from
the OBE interaction and adopt the common cutoff, the reasonable
value is around $\Lambda_i\approx 0.8$ GeV. Then a problem emerges.
Since the $m_\rho/m_\omega\approx0.78$ GeV, their form factors
$F_i(q)=(\Lambda_i^2-m_i^2)/(\Lambda_i^2-q^2)\to 0$, and then the
short-range $\rho$, $\omega$ potentials go to 0. On the other hand,
if we adopt the scaled cutoff scheme, the $\rho$, $\omega$
potentials could be much larger than the $\pi$ potential. It may
also be doubtful since the OPE potential should play a more
important role than the one-vector-exchange (OVE) potentials for the
state very close to the thresholds. To deal with this problem, we
adopt a compromise scheme---assuming
$\Lambda_i=m_i\left[1+\alpha(\Lambda_{\text{QCD}}/m_i)^2\right]$.

Adopting the compromise scheme for the cutoff, we get the results in
the OBE potential case for the $1(1^+)$ system in Table \ref{tab: cc
OBE}. We find a quasibound state, and its behavior is just like the
situation in the OPE potential case.

    \begin{table}[!htb]
        \centering
        \scalebox{0.9}{
            \begin{tabular}{c|ccc}  \hline
                    $\alpha$              & $2.0$ & $2.2$ & $2.4$  \\
                \hline
                    Energy (MeV)          & $-3.98-0.31i$          & $-14.27-0.25i$         & $-31.06$  \\
                    $\sqrt{\langle r^2\rangle}$ (fm)               & $1.7-0.1i$            & $1.0$         & $0.7$    \\
                    $\langle\tilde{\psi}_i|\psi_i\rangle\times 100$             & ($81.6-1.5i$/$1.7$/$14.8+1.4i$  & ($66.5-0.4i$/$1.1$/$28.2+0.4i$ & ($55.5$/$0.7$/$37.3$  \\
                    & /$0.5$/$1.2+0.1i$/$0.2$)   & /$0.8$/$3.1$/$0.2$)  &/$1.1$/$5.3/0.2$) \\
                \hline
            \end{tabular}
        }
        \centering
\caption{\small  Solutions for the double-charm hexaquark with
$I(J^{P})=1(1^+)$ in the OBE potential case with the
$\theta=20^{\circ}$. The energies are given relative to the
threshold of $\Lambda_c\Sigma_c$. The $\sqrt{\langle
r^2\rangle}=[e^{3i\theta}\int_{0}^{\infty}\{\psi(re^{i\theta})\}^2
r^2dr]^{1/2}$ is the root-mean-square (RMS) radius. The
$\langle\tilde{\psi}_i|\psi_i\rangle=e^{i\theta}\int_{0}^{\infty}\{\psi_i(re^{i\theta})\}^2dr$
is the amplitude corresponding to the $i$-th channel of
$\Lambda_c\Sigma_c(^3S_1,^3D_1)$, $\Lambda_c
\Sigma_c^*(^3S_1,^3D_1)$, $\Sigma_c
\Sigma_c(^3S_1,^3D_1)$.}\label{tab: cc OBE}
    \end{table}

We take the $\alpha=2.2$ as an example and show the eigenvalue
distribution in the OBE potential case, as shown in Fig. \ref{fig:
Hcc csm obep}. Obviously, the quasibound state pole is located on
the first Riemann sheets (physical sheets) corresponding to the
$\Lambda_{c}\Sigma_c$, $\Lambda_{c}\Sigma_c^*$ and
$\Sigma_{c}\Sigma_c$ channels and the second Riemann sheets
(unphysical sheets) corresponding to the $\Lambda_c\Lambda_c\pi$
three-body channel. In Fig. \ref{fig: ur obep}, we choose
$\theta=20^\circ$ and plot the real and imaginary parts of its wave
function. Obviously, the $S-$wave $\Lambda_{c}\Sigma_c$ and
$\Lambda_{c}\Sigma_c^*$ channels dominate the state.

    \begin{figure}[htbp]
        \includegraphics[width=240pt]{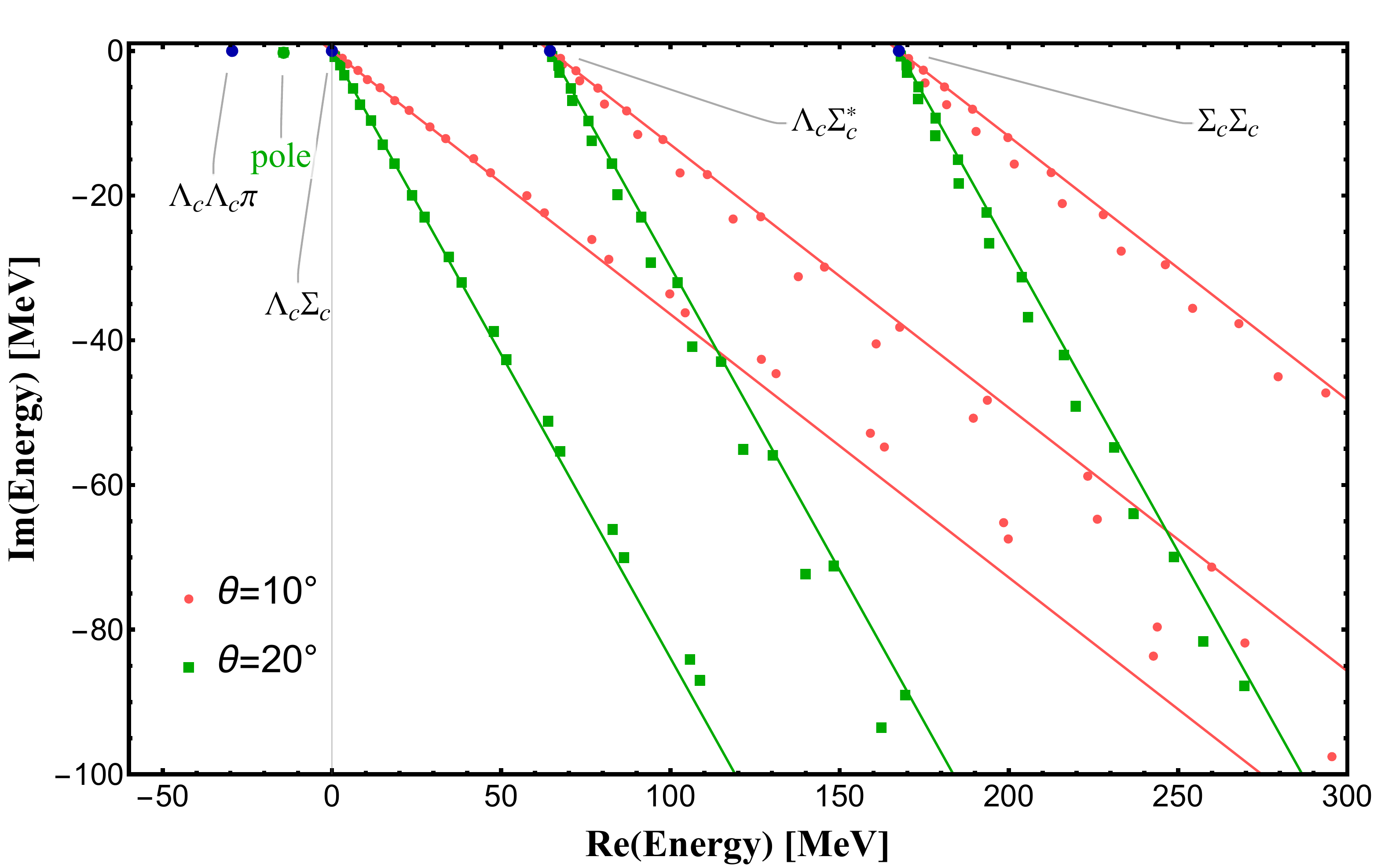}
\caption{The eigenvalue distribution of the double-charm hexaquark
with the $I(J^P)=1(1^+)$. The $\alpha=2.2$ in the OBE potential
case. The red (green) points (square point) and lines correspond to
the situation with the complex rotation angle $\theta=10^{\circ}
(20^{\circ})$.}\label{fig: Hcc csm obep}
    \end{figure}

    \begin{figure}[htbp]
        \subfigure[Re$\big(u_{i}(r)\big)$]{ \label{fig: ur obep re}
            \includegraphics[width=210pt]{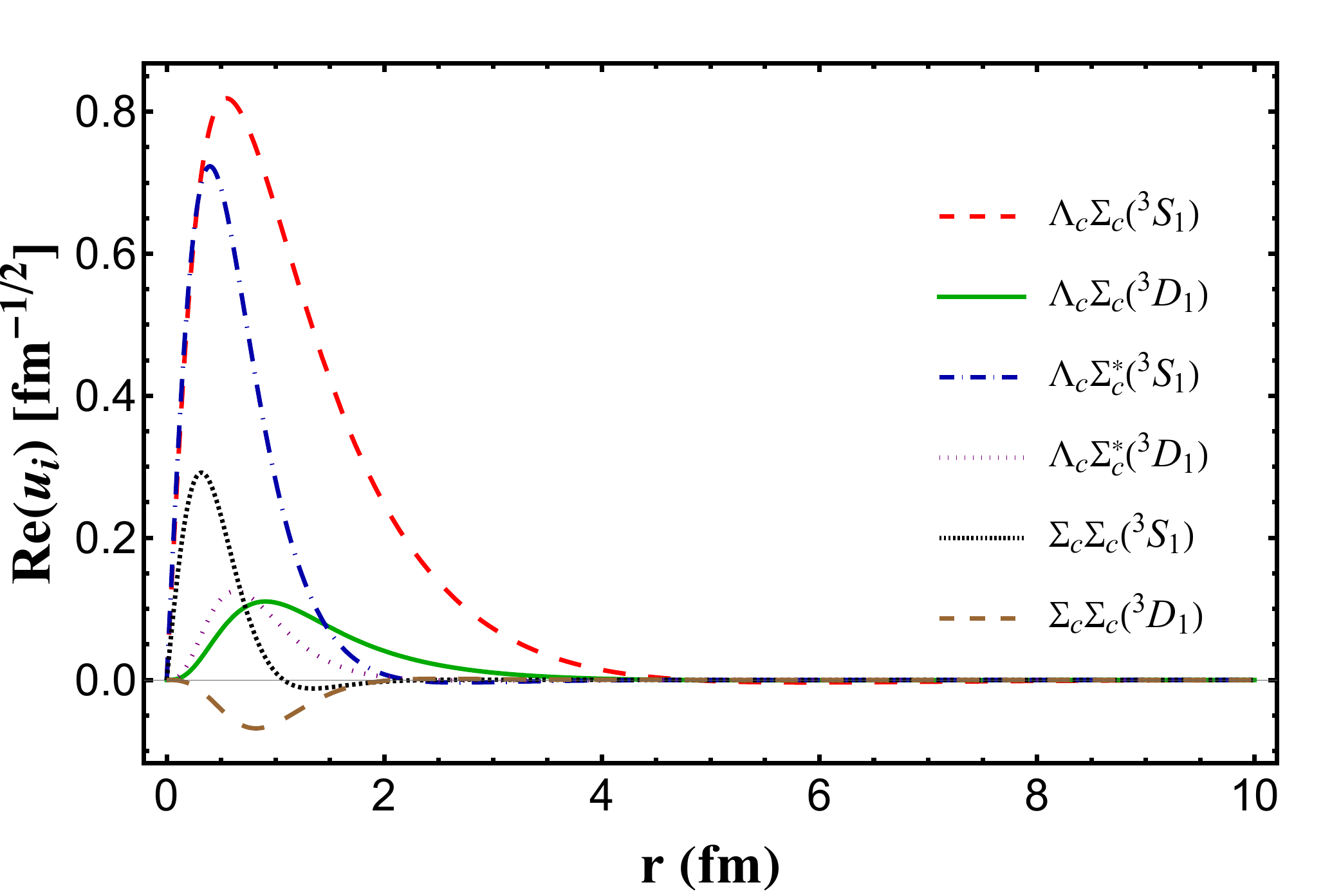}}\hspace{40pt}
        \subfigure[Im$\big(u_{i}(r)\big)$]{ \label{fig: ur obep im}
            \includegraphics[width=210pt]{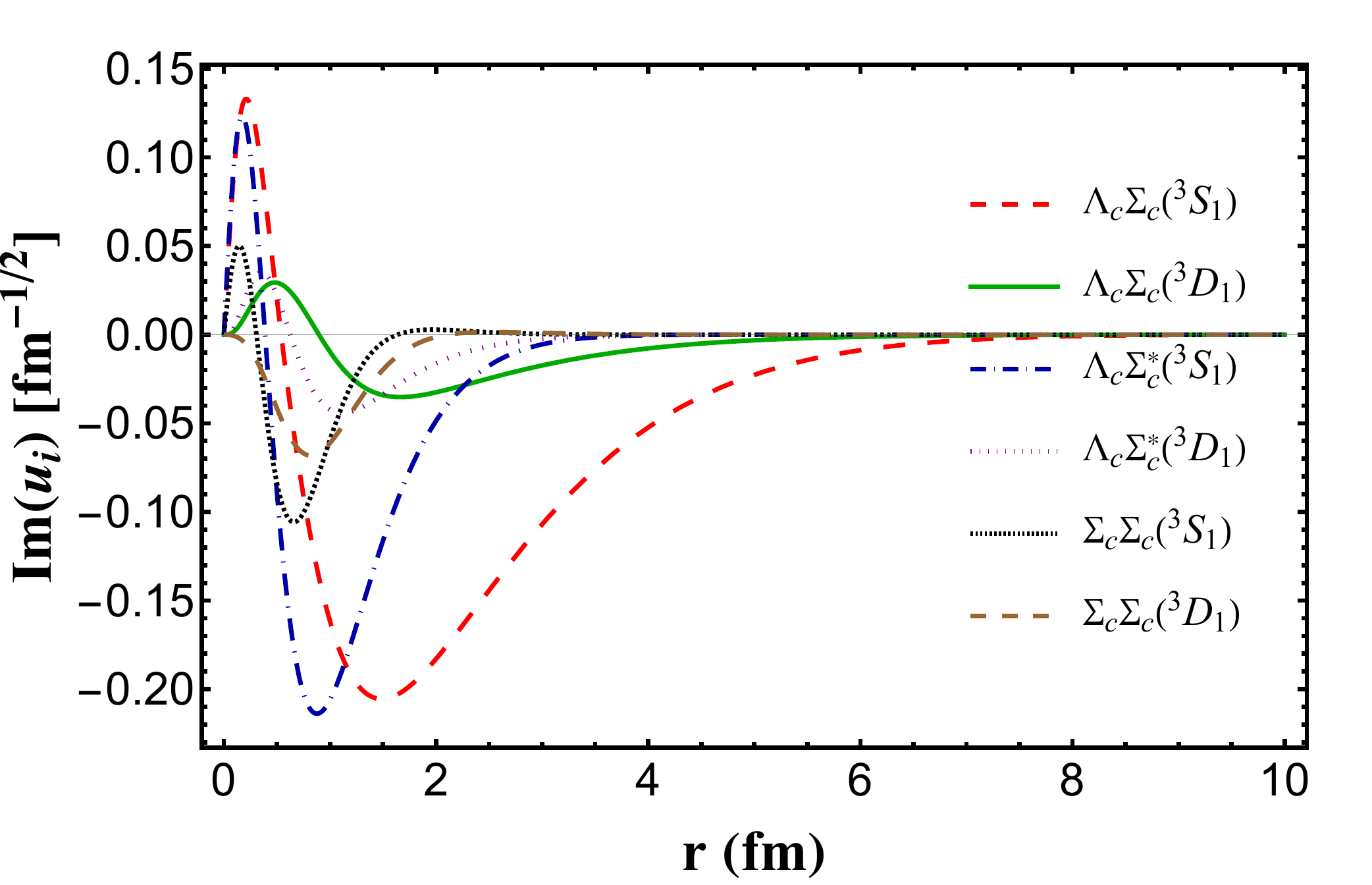}}
\caption{The wave functions $u_i(r)(i=1,2,3,4,5,6)$ of the
double-charm hexaquark with the $I(J^P)=1(1^+)$. The rotation angle
$\theta=20^\circ$ and the parameter $\alpha=2.2$ in the OBE
potential case. The two diagrams correspond to: (a) the real part of
the $u_i(r)$ (b) the imaginary part of the $u_i(r)$.}\label{fig: ur
obep}
    \end{figure}

\subsection{The OBE potential results for the hidden-charm hexaquark system}

In this part, we discuss the molecule system of the hidden-charm
hexaquark with the OBE potential. The relevant potentials are
similar to those of the double-charm cases, and they could be
connected by making a G parity transformation for the propagators.
In other words,
    \begin{eqnarray}
        V^{AB}&=&(-1)^{G_i} V^{A\bar{B}},  \nonumber\\
    \end{eqnarray}
where the $A$, $B$ are the charmed baryons, and $G_i$ is the G
parity of the $i$ propagator, as shown in Table \ref{tab: G parity}.

    \begin{table}[htbp]
    \caption{G parity of the light mesons.  }\label{tab: G parity}
    \begin{tabular}{c|ccccc}\hline\hline
        Meson&$\pi$&$\eta$&$\sigma$&$\rho$&$\omega$\\
        \hline
        G    &-1   &1     &1       &1     &-1\\
        \hline\hline
    \end{tabular}
\end{table}

Considering the multichannel coupling effect, we adopt the channels
in Table \ref{tab: channel} for the systems $1^+(0^{--})$,
$1^-(0^{-+})$, $1^+(1^{--})$ and $1^-(1^{-+})$. Compared with the
double-charm case, the number of the hidden-charm systems doubles
for the existence of the C or G parity number. For the $1^-(0^{-+})$
case, we do not find a pole in a reasonable cutoff region. For the
positive G parity case $1^+(0^{--})$, we find a quasibound state,
and the results are given in Table \ref{tab: obe 0+}. This system
has only one channel, and one may find a clearer width behavior than
in the other coupled-channel systems. When the $\alpha=1.5$, we find
a quasibound state with the binding energy of -5.37 MeV, and the
width of 1.72 MeV. Obviously, the width is very close to the upper
limit of this pole---the width of the $\Sigma_c$.

    \begin{table}[!htb]
        \centering
        \scalebox{0.9}{
            \begin{tabular}{c|ccc}  \hline
                    $\alpha$              & $1.5$ & $1.8$ & $2.1$  \\
                \hline
                 Energy (MeV)          & $-5.37-0.86i$          & $-13.29-0.46i$         & $-25.63-0.05i$  \\
                 $\sqrt{\langle r^2\rangle}$ (fm)                & $1.7$            & $1.2$         & $0.9$    \\
                \hline
            \end{tabular}
        }
        \centering
\caption{\small  Solutions for the hidden-charm hexaquark with
$I^G(J^{PC})=1^+(0^{--})$ in the OBE potential case with the
$\theta=20^{\circ}$. The energies are given relative to the
threshold of $\Lambda_c\bar{\Sigma}_c$. The $\sqrt{\langle
r^2\rangle}=[e^{3i\theta}\int_{0}^{\infty}\{\psi(re^{i\theta})\}^2
r^2dr]^{1/2}$ is the root-mean-square (RMS) radius.}\label{tab: obe
0+}
    \end{table}

For the vector cases, both the $1^+(1^{--})$ and $1^-(1^{-+})$
systems can form a quasibound state. We first discuss the
$1^-(1^{-+})$ system. Similar to the double-charm case $1(1^+)$, the
first diagonal $S-$wave OPE potential
$V_\pi^{\{\Lambda_c\bar{\Sigma}_c\}\to\{\Lambda_c\bar{\Sigma}_c\}}$
is repulsive, and the third diagonal $S-$wave OPE potential
$V_\pi^{[\Lambda_c\bar{\Sigma}_c^*]\to[\Lambda_c\bar{\Sigma}_c^*]}$
is attractive. The numerical results of this pole are given in Table
\ref{tab: obe 1-}, and the main contributions are from the $S-$wave
channels $\{\Lambda_c\bar{\Sigma}_c\}$ and
$[\Lambda_c\bar{\Sigma}_c^*]$. In fact, this quasibound state is
very similar to the pole in the double-charm case $1(1^+)$. They
have very similar results with the same energy, including the
widths, constituents and sizes. For example, taking the
$\alpha=2.03$, we get the binding energy of -4.21 MeV, the width of
0.56 MeV, the RMS of 1.7-0.1$i$ fm and
$\langle\tilde{\psi}_i|\psi_i\rangle\times
100$=(83.8-1.3$i$/2.5/13.2+1.3$i$/0.5), which are very similar to
the results in Table \ref{tab: cc OPE} with $\alpha=2.0$.

    \begin{table}[!htb]
        \centering
        \scalebox{0.9}{
            \begin{tabular}{c|ccc}  \hline
                    $\alpha$              & $2.1$ & $2.3$ & $2.5$  \\
                \hline
                 Energy (MeV)          & $-6.43-0.32i$          & $-15.33-0.23i$         & $-27.88-0.01i$  \\
                 $\sqrt{\langle r^2\rangle}$ (fm)               & $1.5$            & $1.0$         & $0.8$    \\
                 $\langle\tilde{\psi}_i|\psi_i\rangle\times 100$             & ($80.0-1.1i$/$2.5$  & ($69.7-0.4i$/$2.3$ & ($61.2$/$2.2$  \\
                    & /$17.0+1.1i$/$0.6$)   & /$27.1+0.4i$/$0.9$)  &/$35.4$/$1.2$) \\
                \hline
            \end{tabular}
        }
        \centering
\caption{\small Solutions for the hidden-charm hexaquark with
$I^G(J^{PC})=1^-(1^{-+})$ in the OBE potential case with the
$\theta=20^{\circ}$. The energies are given relative to the
threshold of $\Lambda_c\bar{\Sigma}_c$. The $\sqrt{\langle
r^2\rangle}=[e^{3i\theta}\int_{0}^{\infty}\{\psi(re^{i\theta})\}^2
r^2dr]^{1/2}$ is the root-mean-square (RMS) radius. The
$\langle\tilde{\psi}_i|\psi_i\rangle=e^{i\theta}\int_{0}^{\infty}\{\psi_i(re^{i\theta})\}^2dr$
is the amplitude corresponding to the $i$-th channel of
$\{\Lambda_c\bar{\Sigma}_c\}(^3S_1,^3D_1)$,
$[\Lambda_c\bar{\Sigma}_c^*](^3S_1,^3D_1)$.}\label{tab: obe 1-}
    \end{table}

Then, we consider the $1^+(1^{--})$ case. Contrary to the
$1^-(1^{-+})$, the first diagonal $S-$wave OPE potential
$V_\pi^{[\Lambda_c\bar{\Sigma}_c]\to[\Lambda_c\bar{\Sigma}_c]}$ is
attractive, and the third diagonal $S-$wave OPE potential
$V_\pi^{{\Lambda_c\bar{\Sigma}_c^*}\to{\Lambda_c\bar{\Sigma}_c^*}}$
is repulsive. In other words, the first channel could form a bound
or quasibound state alone. After considering the coupled-channel
effect, we can obtain the solutions, as shown in Table \ref{tab: obe
1+}. Different from the other cases, this system has two poles when
$\alpha\gtrsim1.9$. The first one is close to the threshold of the
$\Lambda_c\bar{\Sigma}_c$, and the second to the
$\Lambda_c\bar{\Sigma}_c^*$. These two poles are quite different.

\begin{table}[!htb]
    \centering
    \scalebox{0.9}{
        \begin{tabular}{c|cccc}  \hline
            pole                &   $\alpha$              & $1.7$ & $1.9$ & $2.1$  \\
            \hline
            \multirow{4}{*}{1}  & Energy (MeV)          & $-7.28-0.45i$          & $-15.37-0.23i$         & $-27.66$  \\
            &    $\sqrt{\langle r^2\rangle}$ (fm)               & $1.5$            & $1.1$         & $0.9$    \\
            &    $\langle\tilde{\psi}_i|\psi_i\rangle\times 100$            & ($92.7-0.2i$/$3.0$/$2.9+0.2i$  & ($89.7-0.1i$/$3.5$/$3.8+0.1i$ & ($86.0$/$4.3$/$3.8$  \\
            &   & /$0.$/$0.5$/$0.9$)   & /$0.$/$1.3$/$1.7$)  &/$0.$/$3.0/2.8$) \\
            \hline
            \multirow{4}{*}{2}  & Energy (MeV)          & $$          & $63.55-1.36i$         & $53.95-3.25i$  \\
            &    $\sqrt{\langle r^2\rangle}$ (fm)               & $$            & $2.5-0.7i$         & $1.1-0.2i$    \\
            &    $\langle\tilde{\psi}_i|\psi_i\rangle\times 100$           &   & ($-0.6i$/$-1.6+0.8i$/$91.4-9.9i$ & ($-0.9$/$-0.6+9.9i$/$67.3-9.6i$  \\
            &   &    & /$5.9+8.1i$/$4.1+1.6i$/$0.1+0.1i$)  &/$3.9-1.1i$/$29.3+0.4i$/$1.0+0.3i$) \\
            \hline
            \multirow{4}{*}{2I}  & Energy (MeV)          & $$          & $63.66-1.45i$         & $53.77-2.28i$  \\
            &   $\sqrt{\langle r^2\rangle}$ (fm)               & $$            & $2.3-1.1i$         & $1.1-0.1i$    \\
            &    $\langle\tilde{\psi}_i|\psi_i\rangle\times 100$           &   & ($-0.6-0.5i$/$-1.1+1.5i$/$93.7-3.5i$ & ($0.7i$/$2.0+6.1i$/$72.5-4.9i$  \\
            &   &    & /$2.1+0.4i$/$5.7+2.1i$/$0.1$)  &/$1.6-0.1i$/$23.4-1.8i$/$0.5$) \\
            \hline
        \end{tabular}
    }
    \centering
\caption{\small Solutions for the hidden-charm hexaquark with
$I^G(J^{PC})=1^+(1^{--})$ in the OBE potential case with the
$\theta=20^{\circ}$. The energies are given relative to the
threshold of $\Lambda_c\bar{\Sigma}_c$. The $\sqrt{\langle
r^2\rangle}=[e^{3i\theta}\int_{0}^{\infty}\{\psi(re^{i\theta})\}^2
r^2dr]^{1/2}$ is the root-mean-square (RMS) radius. The
$\langle\tilde{\psi}_i|\psi_i\rangle=e^{i\theta}\int_{0}^{\infty}\{\psi_i(re^{i\theta})\}^2dr$
is the amplitude corresponding to the $i$-th channel of
$[\Lambda_c\bar{\Sigma}_c](^3S_1,^3D_1)$,
$\{\Lambda_c\bar{\Sigma}_c^*\}(^3S_1,^3D_1)$,
$\Sigma_c\bar{\Sigma}_c(^3S_1,^3D_1)$. The data of row "2" are the
results we actually adopt, and the $q_0$ herein is from Table
\ref{tab: q0}. The data of row "2I" are from the instantaneous
approximation with $q_0=0$.}\label{tab: obe 1+}
\end{table}

\begin{figure}[htbp]
    \includegraphics[width=240pt]{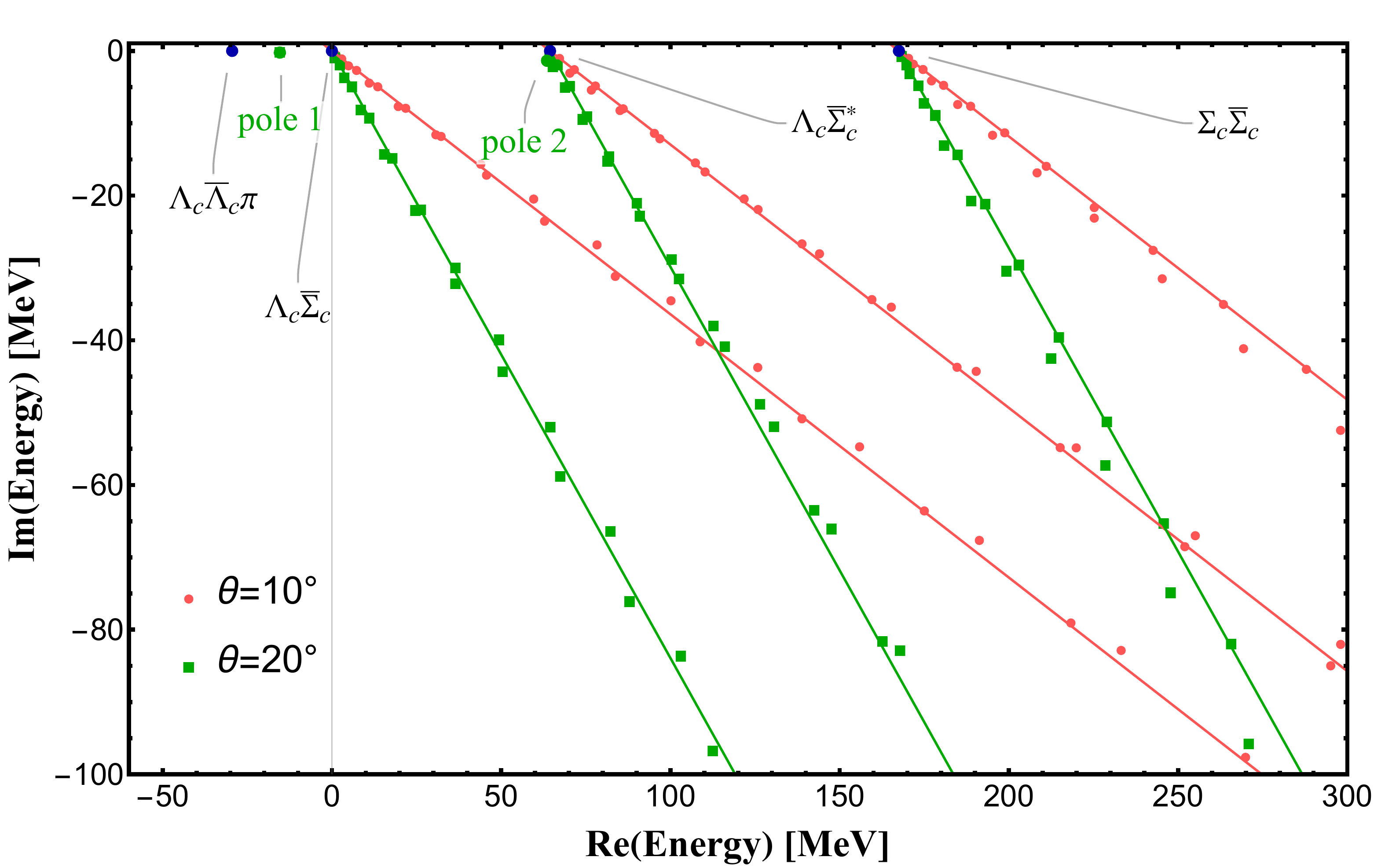}
\caption{The eigenvalue distribution of the hidden-charm hexaquark
with the $I(J^P)=1^+(1^{--})$. The $\alpha=1.9$ in the OBE potential
case. The red (green) points (square point) and lines correspond to
the situation with the complex rotation angle $\theta=10^{\circ}
(20^{\circ})$.}\label{fig: Hccbar csm obep}
\end{figure}

To study their characters, we illustrate the energy distribution
with the $\alpha=1.9$ in Fig. \ref{fig: Hccbar csm obep}. The first
pole mainly consists of the $S$-wave $[\Lambda_c\bar{\Sigma}_c]$
with the energy of $-15.37-0.23i$ MeV. It is located on the first
Riemann sheets (physical sheets) corresponding to the channels
$[\Lambda_{c}\bar{\Sigma}_c]$, $\{\Lambda_{c}\bar{\Sigma}_c^*\}$ and
$\Sigma_{c}\bar{\Sigma}_c$, and the second Riemann sheets
(unphysical sheets) corresponding to the three-body channel
$\Lambda_c\bar{\Lambda}_c\pi$ . The second pole mainly consists of
the $S$-wave $\{\Lambda_c\bar{\Sigma}_c^*\}$ with the energy of
$63.55-1.36i$ MeV relative to the threshold of
$[\Lambda_c\bar{\Sigma}_c]$. It is located on the first Riemann
sheets (physical sheets) corresponding to the channels
$\{\Lambda_{c}\bar{\Sigma}_c^*\}$ and $\Sigma_{c}\bar{\Sigma}_c$,
and the second Riemann sheets (unphysical sheets) corresponding to
the $[\Lambda_{c}\bar{\Sigma}_c]$ and the
$\Lambda_c\bar{\Lambda}_c\pi$ three-body channel. Obviously, the
second pole is a Feshbach-type resonance---if we turn off the
$\Lambda_{c}\bar{\Sigma}_c^*$ channels, it disappears. In addition,
the width of the first pole is totally from the three-body decay
process. However, the width of the second one may have additional
sources.

To figure out this source, we take the instantaneous approximation
$q_0=0$. We also obtain two similar poles that correspond to poles
"1" and "2" in Table \ref{tab: obe 1+}. The one corresponding to
pole "1" turns into a bound state, and we have discussed this case
in Sec. \ref{sec:OPE result}. We will focus on the other pole, whose
numerical results are listed in row "2I" of Table \ref{tab: obe 1+}. Its energy
becomes $63.66-1.45i$ MeV for the $\alpha=1.9$ case, whose real part
is nearly unchanged compared with "2" pole with energy of
$63.55-1.36i$ MeV. However, its width changes and does not disappear
like the other quasibound state. As discussed in the Sec.
\ref{sec:OPE result}, the width from the three-body decay will
vanish under the instantaneous approximation. Therefore, the width
in "2I" is totally from the two-body decay process, such as the
decay of the pole $\text{"2I"}\to[\Lambda_c\bar{\Sigma}_c]$. We
infer from the change of the width that the two-body decay plays a
more important role than the three-body decay in this resonance.

\begin{figure}[htbp]
    \subfigure[Re$\big(u_{i}(r)\big)$]{ \label{fig: ur obep ccbar re}
        \includegraphics[width=210pt]{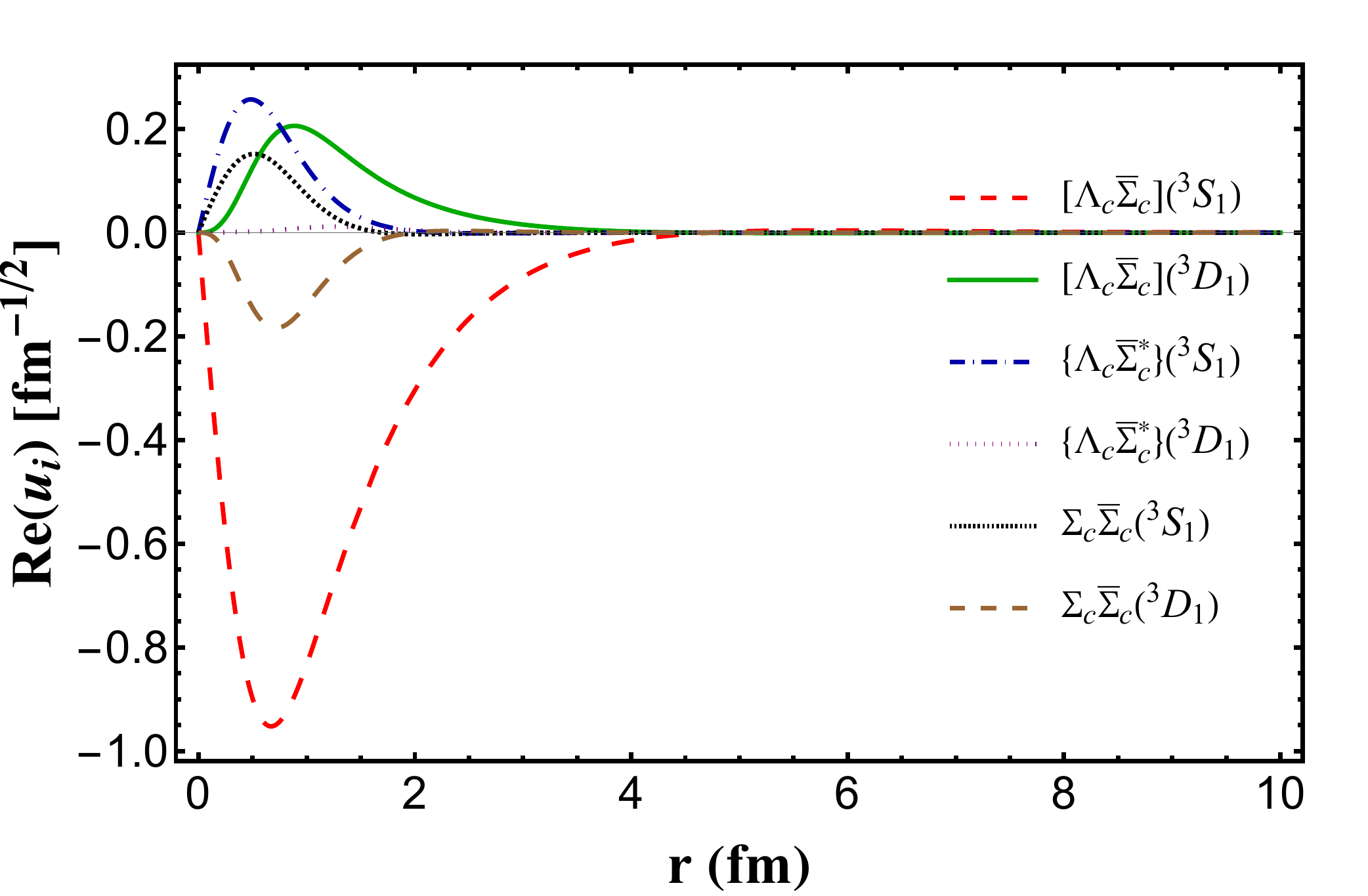}}\hspace{40pt}
    \subfigure[Im$\big(u_{i}(r)\big)$]{ \label{fig: ur obep ccbar im}
        \includegraphics[width=210pt]{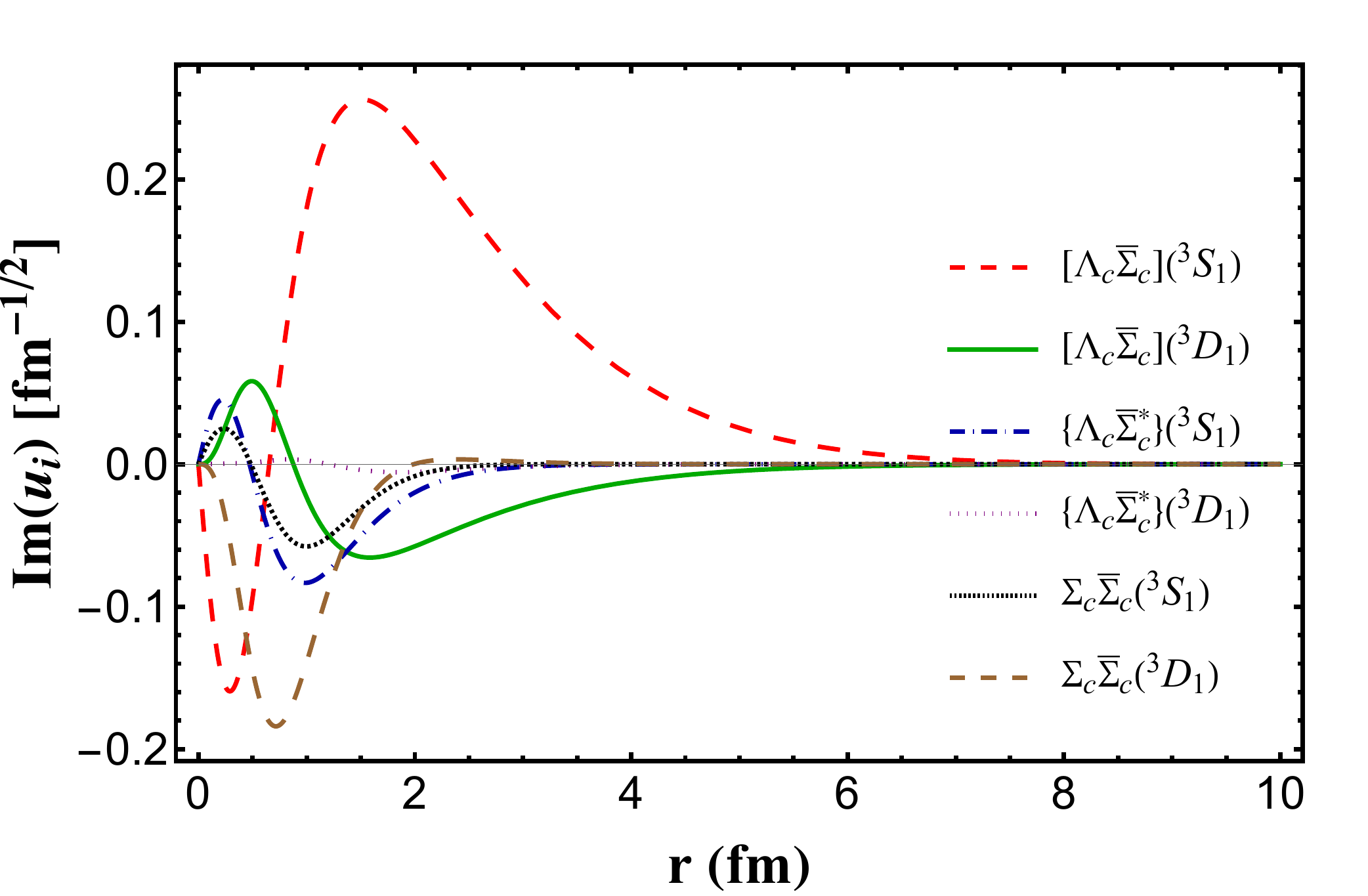}}
    \subfigure[Re$\big(u_{i}(r)\big)$]{ \label{fig: ur obep ccbar re2}
    \includegraphics[width=210pt]{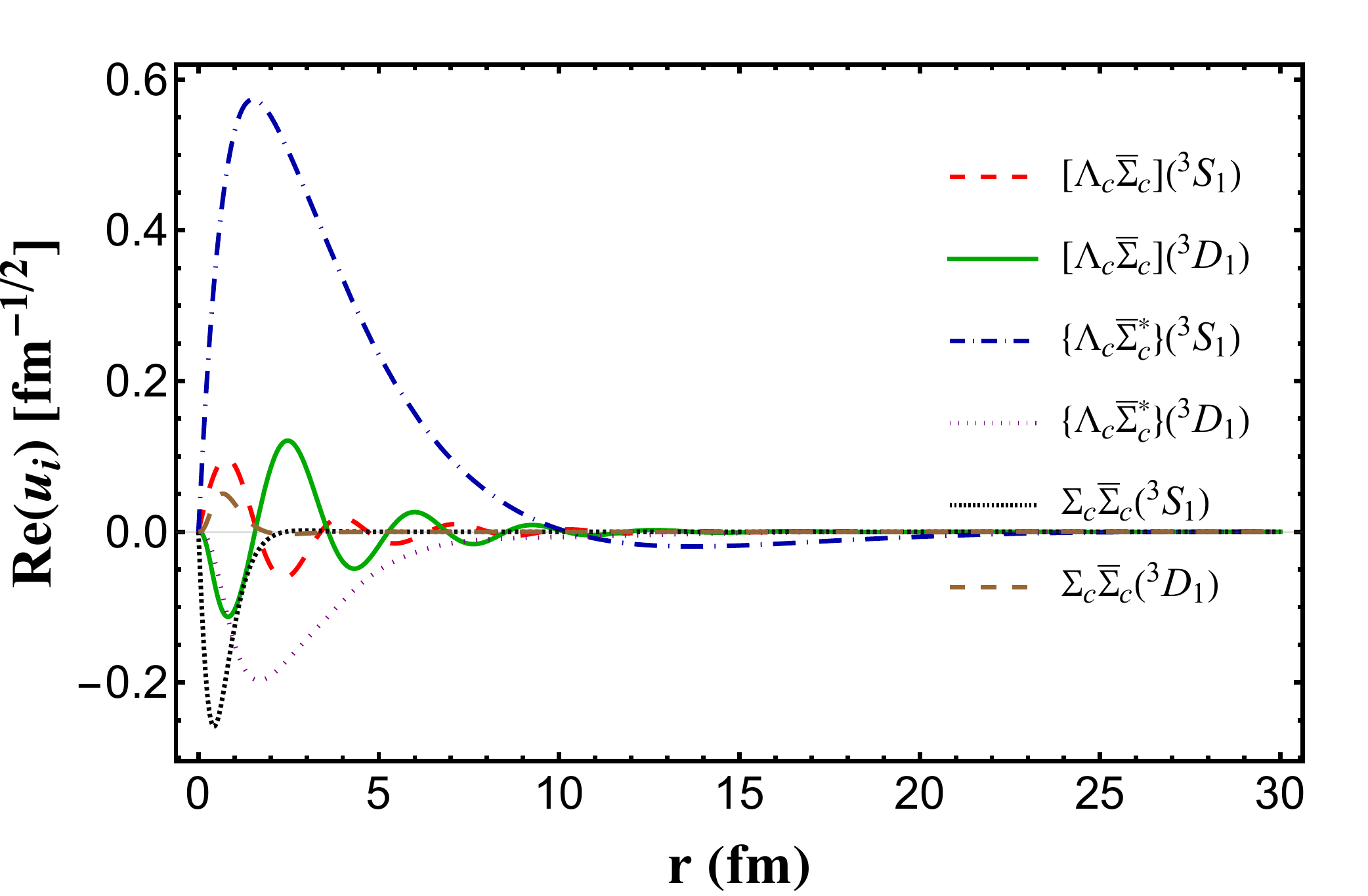}}\hspace{40pt}
    \subfigure[Im$\big(u_{i}(r)\big)$]{ \label{fig: ur obep ccbar im2}
    \includegraphics[width=210pt]{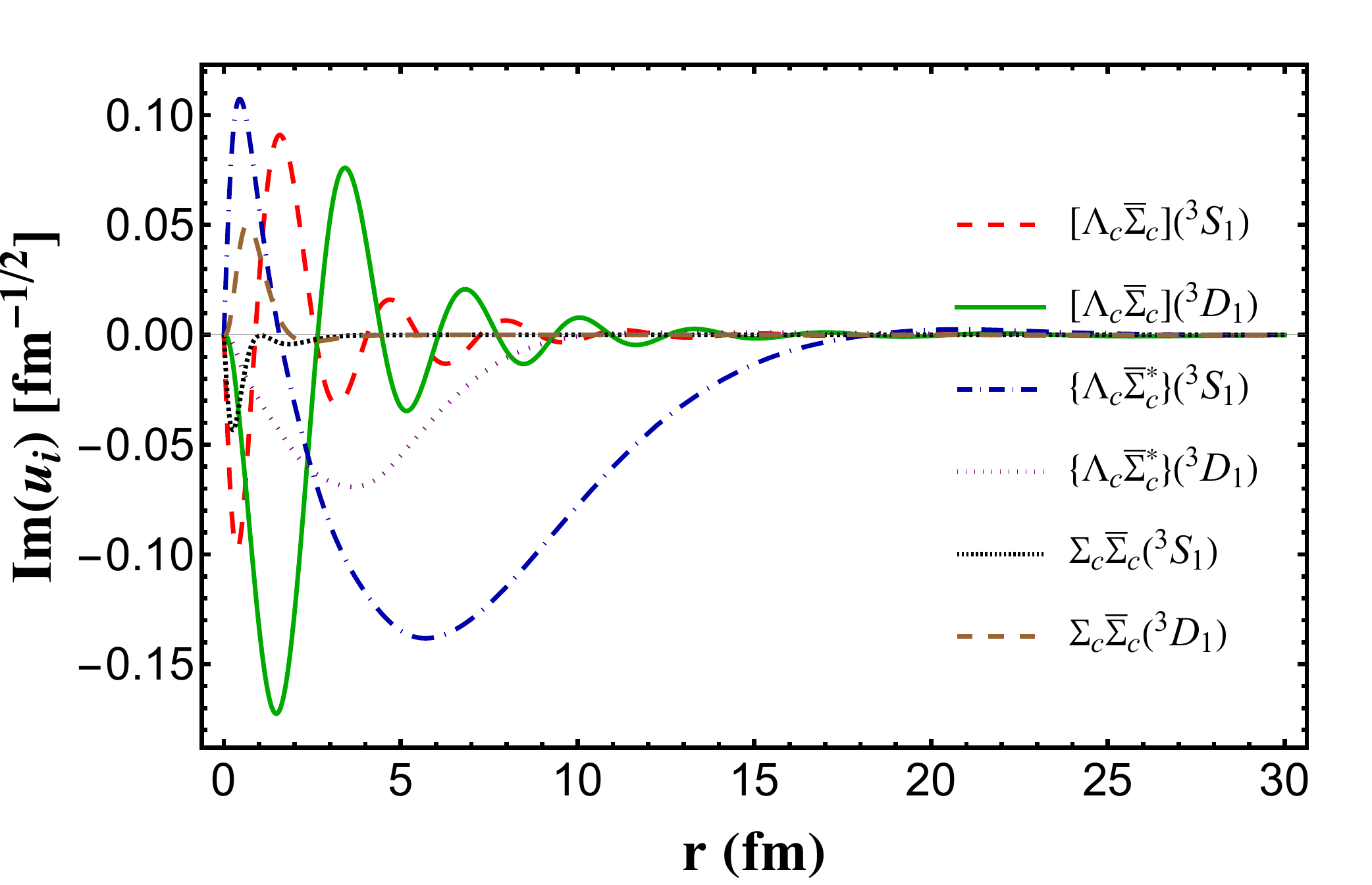}}
\caption{The wave functions $u_i(r)(i=1,2,3,4,5,6)$ of the
hidden-charm hexaquark with the $I(J^P)=1^+(1^{--})$. The rotation
angle $\theta=20^\circ$ and the parameter $\alpha=1.9$ in the OBE
potential case. The four diagrams correspond to: (a) the real part
of the $u_i(r)$ for the first pole (b) the imaginary part of the
$u_i(r)$ for the first pole (c) the real part of the $u_i(r)$ for
the second pole (d) the imaginary part of the $u_i(r)$ for the
second pole.}\label{fig: ur obep ccbar}
\end{figure}

Finally, we also plot the wave functions of poles "1" and "2", as
shown in Fig. \ref{fig: ur obep ccbar}. We choose $\theta=20^\circ$
and present the real and imaginary parts of their wave functions.
Obviously, the $S-$wave $[\Lambda_{c}\bar{\Sigma}_c]$ and
$\{\Lambda_{c}\bar{\Sigma}_c^*\}$ channels dominate the first pole,
and $S-$wave $\{\Lambda_{c}\bar{\Sigma}_c^*\}$ channel dominate the
second pole.

\section{Summary}\label{sec:summary}

In this work, we use the complex scaling method to study the
double-charm and hidden-charm hexaquark states in the molecule
picture. In order to include the coupled-channel effects, we
consider the channels $\Lambda_c\Sigma_c^{(*)}$ (or
$\Lambda_c\bar{\Sigma}_c^{(*)}$) and $\Sigma_c\Sigma_c$ (or
$\Sigma_c\bar{\Sigma}_c$). We also take into account the $S-D$ wave
mixing effect in this deuteronlike dibaryon (hidden-charm
baryonium), as shown in Table \ref{tab: channel}.

We adopt the effective Lagrangians constructed in terms of the heavy
quark symmetry and chiral symmetry. To figure out the influence of
the long-range pion exchange in the formation of the bound states
and resonances, we adopt the OPE potential for the double-charm
hexaquark system. And we also give the numerical results with the
OBE potential for the double-charm and hidden-charm hexaquark
systems.

The OPE potentials of the $\Lambda_c\Sigma_c^{(*)}$ and the
tetraquark $DD^{*}$ systems are similar. They both have an imaginary
part. This imaginary part comes from the processes
$\Sigma_c^{(*)}\to \Lambda_c\pi$ ($D^{*}\to D\pi$), which can be
naturally understood in the framework of the CSM. In the study of
the double-charm hexaquark system with the OPE potential, we find a
quasibound state in the $1(1^+)$ system, which mainly consists of
the $S$-wave $\Lambda_c\Sigma_c$ and $\Lambda_c\Sigma_c^*$. When
taking $\Lambda_\pi=1$ GeV, the binding energy relative to the
$\Lambda_c\Sigma_c$ is -5.6 MeV, and the width is 0.86 MeV. As
explained in Sec. \ref{sec:OPE result}, this width is totally from
the $\Lambda_c\Lambda_c\pi$ three-body decay process. For the system
with $1(0^+)$, we do not find a bound state or resonance.

We also employ the OBE potential to include the medium- and
short-range interactions. And we get a similar result compared with
the OPE case---only one pole is found. Its binding energy relative
to the $\Lambda_c\Sigma_c$ and width are -14.27 MeV and 0.50 MeV
respectively when taking the $\alpha=2.2$.  The $S$-wave
$\Lambda_c\Sigma_c$ and $\Lambda_c\Sigma_c^*$ are the dominant
constituents, and the $D$-wave constituents still provide small
contributions.

For the hidden-charm hexaquark systems, we find more poles. In the
$1^-(0^{-+})$ case, we do not find a pole in a reasonable cutoff
region. However, we find a quasibound state in the $1^+(0^{--})$
case with a single channel. When $\alpha=1.5$, the binding energy is
-5.37 MeV relative to the $\Lambda_c\bar{\Sigma}_c$ threshold. Its
width is 1.72 MeV which is very close to the width of $\Sigma_c$. In
the vector cases, we find poles in both the $1^+(1^{--})$ and
$1^-(1^{-+})$ cases. For the $1^-(1^{-+})$ case, we find a
quasibound state, which behaves just like the mentioned pole in the
$1(1^+)$ double-charm hexaquark. We get this pole in similar region
$\alpha\in [2.0\sim2.5]$. For the same energy or cutoff, their
widths, sizes and constituents are close to each other too. For the
$1^+(1^{--})$ case, we find two poles---a pole close to the
$\Lambda_c\bar{\Sigma}_c$ threshold and the other close to the
$\Lambda_c\bar{\Sigma}_c^*$ threshold. Taking $\alpha=1.9$, the
first pole as a quasibound state has a binding energy of -15.37 MeV
and a width of 0.46 MeV, and the $S$-wave
$[\Lambda_c\bar{\Sigma}_c]$ plays a dominant role. The second pole
is a resonance, whose energy relative to the
$\Lambda_c\bar{\Sigma}_c$ threshold is $63.55-1.36i$ MeV. Different
from the above quasibound states whose widths are totally from the
three-body decay, its width arises from two sources---the three-body
decay and the two-body decay. As shown in Table \ref{tab: obe 1+},
its width does not vanish when we get rid of the three-body decay
effect, and the contribution from the two-body decay is apparently
larger. We also plot the wave functions of these two poles, see in
Fig. \ref{fig: ur obep ccbar}.

In summary, we have found some quasibound states and resonances in
these systems. One could look for these states through their strong
decay patterns, such as $[\Lambda_c\bar{\Sigma}_c]$ and
$\Lambda_c\bar{\Lambda}_c\pi$ invariant mass distributions in the
hidden-charm baryonium and $\Lambda_c\Lambda_c\pi$ invariant mass
distribution in the double-charm dibaryon. Especially, the
$1^+(0^{--})$ and $1^-(1^{-+})$ hidden-charm hexaquark molecular
states are very interesting. These isovector mesons have exotic
$J^{PC}$ quantum numbers which are not accessible to the
conventional $q\bar q$ mesons. Hopefully, this work could be helpful
for future experimental search of the hexaquark states at facilities
such as LHCb and BelleII.

\acknowledgments{This research is supported by the National Science
Foundation of China under Grants No. 11975033, No. 12070131001 and
No. 12147168.}

    \bibliographystyle{apsrev4-2}
    \bibliography{tot.bib}

\end{document}